\documentclass{aa}

\usepackage[colorlinks=true, linkcolor=blue, citecolor=blue, filecolor=blue, urlcolor=blue, breaklinks=true]{hyperref}

\usepackage{lastpage}
\usepackage{textcomp}
\usepackage{gensymb}
\usepackage{xcolor}
\usepackage{lipsum}
\usepackage{graphicx}
\usepackage{placeins}

\usepackage{natbib,twoopt}
\bibpunct{(}{)}{;}{a}{}{,}  

\usepackage{txfonts}

\begin{document}

\nolinenumbers

   \title{Impact of a binary companion in AGB outflows on CO spectral lines}

   \author{O.~Vermeulen\inst{1}, M.~Esseldeurs\inst{1}, J.~Malfait\inst{1}, T.~Ceulemans\inst{1}, L.~Siess\inst{2}, K.~Matsumoto\inst{3},  F.~De Ceuster\inst{4,1}, T.~Danilovich\inst{5,1}, C.~Landri\inst{1}, \and L.~Decin\inst{1} }

   \institute{Institute of Astronomy, KU Leuven, Celestijnenlaan 200D, 3001 Leuven, Belgium\\
   \email{owen.vermeulen@kuleuven.be}
         \and
    Institut d’Astronomie et d’Astrophysique, Université Libre de Bruxelles (ULB), CP 226, 1050 Brussels, Belgium 
     \and
    Sterrenkundig Observatorium, Universiteit Gent, Krijgslaan 281 S9, B-9000 Gent, Belgium 
     \and 
     Leuven Gravity Institute (LGI), KU Leuven, Celestijnenlaan 200D, 3001 Leuven
     \and
     School of Physics \& Astronomy, Monash University, Wellington Road, Clayton 3800, Victoria, Australia}

   \date{Received / Accepted}

 \abstract  
   {In the late stage of their evolution, low- to intermediate-mass stars pass through the asymptotic giant branch (AGB) phase. AGB stars are characterised  by strong mass loss through dust driven winds. High angular resolution interferometric observations reveal that these winds harbour strong deviations from spherical symmetry, such as spirals and arcs, believed to be caused by hidden (sub-)stellar companions. Such observations are scarce, and  much more often observed features from these systems are molecular spectral lines, where the presence of a companion is less clear.}
   {Our aim is to study the impact of a binary companion on low-$J$ CO spectral lines of AGB star outflows. By varying the orbital separation and wind velocity, we aim to find line shapes characteristic of more complex binary-induced morphologies. This would allow us to quantify whether a companion can be identified from spectral lines, or if it remains hidden in the line profiles.}
   {We generated a new grid of nine 3D models of a mass-losing AGB star using the smoothed particle hydrodynamics (SPH) code \textsc{Phantom}, with three values for both the outflow velocity and orbital separation. We created a novel method for calculating the CO photodissociation in asymmetric outflows in order to determine the size of the emitting envelope. Utilising the 3D non-local thermodynamic equilibrium (NLTE) radiative transfer code \textsc{Magritte}, we created synthetic spectral lines for the low rotational transitions of CO at different inclinations and position angles.}
   {Our hydrodynamical simulations show a variety of morphologies, always with a pronounced spiral structure arising in the orbital plane, but with varying shapes in the meridional plane, and different degrees of global flattening. CO photodissociation closely follows the global morphology, thereby creating a non-spherical emitting region. We find that the low-$J$ CO line profiles can deviate strongly from the parabolic or flat-topped profiles expected from non-resolved spherically symmetric outflows. A variety of line shapes emerge, with two peaks near the terminal velocity, and a central bump near the central velocity being the most pronounced. The line shapes strongly depend on the underlying morphology and inclination. In specific cases the spectral lines can appear parabolic, hiding the presence of a binary companion. }
   {We find that a binary companion can have a pronounced impact on the resulting CO spectral lines, and thus the molecular line profiles can serve as a binary diagnostic. However, the influence of the companion on the line can also go easily unnoticed as the characteristic features can be concealed by the beam profile and the noise of the observations. Therefore, it is easy to misclassify systems as single stars. Neglecting the impact of a binary companion  when modelling the CO spectral lines can thus cause systematic errors on the derived mass-loss rates.  }

    \authorrunning{O. Vermeulen et al.}
    \titlerunning{Impact of binary companion on CO spectral lines}

   \maketitle
\nolinenumbers

\section{Introduction}
\label{section:introduction}
Low- to intermediate-mass stars are stars with typical masses ranging between $\sim0.8$ and $\sim8$ M$_\odot$. After the depletion of hydrogen and helium in their cores, these stars enter the asymptotic giant branch (AGB) phase. In this stage of their evolution, stars are characterised by an extended dusty envelope, and a stellar wind with typical mass-loss rates ranging between 10$^{-8}$ to 10$^{-5}$ M$_\odot$ yr$^{-1}$, and terminal wind velocities between 5 and 30 km s$^{-1}$ \citep{hoffner}. The mechanism driving these winds is generally believed to be a combination of pulsations and radiation pressure on dust grains. Pulsations levitate material  high up in the circumstellar envelope, where the cooler temperature allows the formation of dust grains that can efficiently absorb the infrared stellar radiation, thereby accelerating  outwards and dragging along the surrounding gas \citep{Lamers_Cassinelli_1999}. Depending on the mass, stars in the AGB phase can also experience third dredge-up events, where the outer convective envelope penetrates deeply into the star, enriching the envelope with material originating from nuclear burning and the s-process \citep{1952S,1965ApJ...142..855S, kip}. All together, AGB stars contribute $\sim85\%$ of all the gas and $\sim35\%$ of all the dust to the interstellar medium \citep{tielensISM}. Galactic chemical evolution models predict that AGB stars are the main source of carbon and nitrogen, and about half of the elements heavier than iron in the Universe \citep{enrich}. 

The high-resolution  observations of these outflows, obtained with instruments such as the Atacama Large Millimeter/Submillimeter Array (ALMA) interferometer, reveal a variety of complex morphologies that contrast with the long-held assumption of spherical symmetry \citep[see e.g.][]{Maercker_2016, Kervella_2016, Epaqr, Decin_2020}. The leading hypothesis to explain these structures is a previously unrecognised binary companion that shapes the outflow as the mass-losing AGB star and its companion orbit around the common centre of mass (CoM), as first numerically predicted by \citet{theuns}. The post-AGB stars are also known to harbour complex shapes in their circumstellar envelopes (CSEs), including bipolar nebulae, shells, consecutive rings, and arcs. Similarly, these structures have been linked to the presence of companions \citep{Hans2003, bollen}. Hence, the dominant shaping mechanism of the CSE is believed to be the gravitational interaction of the AGB wind with a (sub\mbox{-})stellar binary companion orbiting around the AGB star \citep{Decin_2020}. This assumption is supported by population synthesis since the binarity fraction of AGB progenitors is found to be above $\sim50\%$, even reaching $\sim100\%$ when planetary companions are also included in the binarity rate \citep{moe_di_stefano, kepler}.

Unfortunately, high angular resolution interferometric observations, where the impact of a companion is readily observable, remain scarce, and are available for only $\sim20$ systems. More frequently, studies of AGB outflows are limited to the emission lines from molecules present in these outflows. Over 100 molecules have been detected in the outflows of AGB stars \citep{molecules1, Daedalean}, which is roughly 50\% of all molecules detected in outer space. The most-studied molecule is CO, whose lower energetic rotational states are easily excited over much of the CSE. By studying these lower transitions, we can probe the entire wind and through modelling, we can derive the mass-loss rate \citep[see e.g.][]{CO1, CO2, 2006, elvire}. In this modelling procedure, the assumption of a single star is always made. Given the likely presence of companions, it is possible to make systematic errors on the derived parameters, especially the mass-loss rate \citep{Decin_2019}. Hence, it is crucial that we model the CO emission lines with the inclusion of the binary companion.

The first steps towards this goal have already been undertaken in \citet{Ward2, Ward1} utilising simpler analytical models to study the effect of different morphologies. Moreover, \citet{Kim_2019}  made use of hydrodynamical simulations, but did not consider detailed radiative transfer to create synthetic spectral lines. Instead, they considered the column density within velocity bins as a proxy for the spectral line. In contrast, here we  use 3D hydrodynamical simulations combined with full non-local thermodynamic equilibrium (NLTE) radiative transfer, thereby including more adequate structures and physics in our models. Our goal is to conduct an initial investigation of how various morphologies arising from these hydro-simulations translate to spectral lines.

Three-dimensional hydrodynamic studies of binary AGB systems have been performed using both grid-based and smoothed particle hydrodynamics (SPH) codes. The early studies of \citet{theuns} and later \citet{MM1, MM2}, already showed that Archimedes-like spiral structures, elongated morphologies, and accretion discs  can form depending on the orbital and wind parameters. Since then, higher resolution simulations became possible, facilitating more comprehensive studies of the formation of different structures, including spirals, arc- and ring-shaped morphologies, equatorial density enhancements, and accretion discs, and the effects of these structures on the mass transfer and orbital evolution \citep[see e.g.][]{Chen_2017, Saladino_2018, Saladino_2019a, Saladino_2019b, Chen_2020, El, SM1, JM1, 2022ApJ...931..142L, aydi_mohamed, JM2, JM3}.  

This paper is outlined as follows. In Sect.~\ref{section:method}, we describe the numerical set-up and parameter space of the HD simulations, outline the radiative transfer to obtain the spectral lines, and describe how we take into account CO photodissociation. Next, in Sect.~\ref{section:results}, we present the hydro-simulations and the resulting spectral lines for the grid of models. In Sect.~\ref{section:discussion}, we discuss the implications and accuracy of our results and outline the necessary steps for the future. We conclude in Sect.~\ref{section:conclusion}.

\section{Methodology}
\label{section:method}
\subsection{Hydrodynamic simulations}
\label{method:HD} 
We use the open-source smoothed particle hydrodynamics (SPH) code \textsc{Phantom}\footnote{\url{https://phantomsph.github.io/}}  \citep{phantom} to model the outflows of AGB stars. SPH codes solve the hydrodynamic equations using a Lagrangian method. These equations are given by 
\begin{align}
    \frac{\mathrm{d} \rho}{\mathrm{d}t} &= - \rho \left( \boldsymbol{\nabla} \cdot \boldsymbol{v} \right), && \text{(continuity equation)} \\
    \frac{\mathrm{d} \boldsymbol{v}}{\mathrm{d}t} &= - \boldsymbol{\nabla} \left( \frac{P}{\rho}\right) + \boldsymbol{a}, && \text{(conservation of momentum)} \\
    \frac{\mathrm{d} u}{\mathrm{d} t} &= -\frac{P}{\rho} (\boldsymbol{\nabla} \cdot \boldsymbol{v}) + \Lambda, && \text{(conservation of energy)} \label{eq:energy}
\end{align}
where $\rho$ is the gas density, $\boldsymbol{v}$ the velocity, $P$ the pressure, $u$ the internal energy, $\boldsymbol{a}$ represents all accelerations from external forces, and $\Lambda$ denotes additional energy losses or gains, which in the present case includes shock heating and H\textsc{i} cooling.

To close this system of differential equations, we assume a polytropic equation of state (EOS), given by  
\begin{equation}
    \label{eq:adiabatic}
    P = (\gamma - 1)\rho u 
\end{equation}
with $\gamma$ the polytropic index. The gas temperature $T$ can be obtained by using Eq.~\ref{eq:adiabatic} and the ideal gas law
\begin{equation}
    \label{eq:ideal}
    P = \frac{\rho k_B T}{\mu m_\mathrm{H}},
\end{equation}
where $k_B$ is Boltzmann's constant, $m_\mathrm{H}$ the mass of a hydrogen atom, and $\mu$ the mean molecular weight. In previous studies, the polytropic index was fixed to a value of 1.2 \citep{SM1}, and $\mu$ was taken to be constant at a value set to 2.38 or 1.26 to mimic either a molecular or atomic gas, respectively \citep{JM2}. To make our simulations more physically consistent, we calculate the values for $\gamma$ and $\mu$  as done in \citet{siess}. For the calculation of the mean molecular weight, we take into account only the most abundant species (H, He, and H$_2$), giving 
\begin{equation}
\label{eq:mu}
    \mu = \frac{(1 + 4 \epsilon_{\mathrm{He}}) n_{\langle \mathrm{H} \rangle } k_B T_g  }{ P_\mathrm{H} + P_{\mathrm{H}_2} + \epsilon_\mathrm{He}n_{\langle \mathrm{H} \rangle } k_B T_g }
\end{equation}
and for the adiabatic index 
\begin{equation}
    \gamma =\frac{ 5P_{\mathrm{H}} + 7 P_{\mathrm{H}_2} + 5\epsilon_\mathrm{He}n_{\langle \mathrm{H} \rangle } k_B T_g  }{ 3P_{\mathrm{H}} + 5P_{\mathrm{H}_2} + 3\epsilon_\mathrm{He}n_{\langle \mathrm{H} \rangle } k_B T_g  }.
\end{equation}
In these expressions, $n_{\langle \mathrm{H} \rangle }$ is the total number of hydrogen atoms per unit volume in the medium, $\epsilon_\mathrm{He}$ is the total number of helium atoms per hydrogen atom which is taken to be $1.04 \times  10^{-1}$. $P_\mathrm{H}$ and $ P_{\mathrm{H}_2}$ represent the partial pressures of atomic and molecular hydrogen, respectively. As for the calculation of $\mu$, contributions from other molecules are neglected because they contribute  less than 1\% of the total mass fraction \citep{siess}. This more physical treatment of the mean molecular weight and adiabatic index impacts the heating and cooling throughout the simulation, which is needed to generate more realistic temperatures.

Additional cooling is implemented in the form of H\textsc{i} cooling, as done in \citet{JM2}. This is calculated using the relation 
\begin{equation}
    \Lambda_{\mathrm{cool\_ H\textsc{i}}} = 7.3 \times 10^{-19} \mathrm{g} \ \mathrm{cm}^3 \times \frac{ n_e n_\mathrm{H} e^{-118 400 \mathrm{K} / T }}{ \rho } \mathrm{erg} \ \mathrm{g}^{-1} \mathrm{s}^{-1},
\end{equation}
where $n_e$ and $n_\mathrm{H}$ are the electron and total hydrogen number densities \citep{Spitzer1978}. This cooling becomes effective at temperatures greater than 3000~K, and is dominant at temperatures higher than 8000~K. As shown in \citet{JM2}, this term is necessary to deal with the high temperatures generated in the high-pressure regions around the companion stars, leading to unphysical instabilities and energetic outflows if not included, as present in \citet{JM1}. This term, together with the adiabatic work due to compression and expansion, and shock heating $\Lambda_{\mathrm{shock}}$ \citep{phantom} are included in Eq.~\ref{eq:energy}.

Both the primary and secondary star are modelled as sink particles. The primary and secondary have an accretion radius $R_{ \mathrm{accr} }$, beyond which particles are accreted if their energy is low enough \citep{phantom}. The actual wind is driven in a free-wind approximation, where the gravity of the primary AGB star is artificially balanced by the radiation force \citep{theuns}. To drive the wind, SPH particles are distributed on spheres around the primary sink particle, and injected into the simulation with a fixed initial velocity \citep[see][]{siess}. 

\subsection{Grid set-up}
\label{section:gridsetup}
We computed a grid of nine models with varying injection velocities and orbital separations. In previous studies \citep[e.g.][]{JM2}, the wind was launched from the stellar surface. However, the wind is launched at the location where dust grains can start forming, i.e. the dust condensation radius. Injecting wind particles from the stellar surface gives rise to unrealistic temperature profiles in the inner region of the simulation. To estimate the wind injection radius, we assume a simple power law for the temperature starting from the surface of the star \citep{2006}. This is given by 
\begin{equation}
\label{eq:Tr}
    T (r)  = T_\star \left( \frac{R_\star}{r} \right)^\zeta
\end{equation}
with $\zeta \approx 0.5$ for optically thin absorption and emission, and $T_\star$ and $R_\star$ the temperature and radius of the star, respectively. We choose values for $T_\star$ and $R_\star$ of 2750~K and 1.267~au, respectively. These values are selected arbitrarily, and are in line with typical values of AGB stars \citep{habing}. Assuming a dust condensation temperature of 1500~K \citep{dustgala}, the dust condensation radius is located at $R_d \approx 4.26 $ au, which is where we inject the free wind. 

In the current paper, we limit ourselves to a low mass-loss rate of $10^{-7}$ M$_\odot$ yr$^{-1}$ for all our models. There are two main reasons for this. First of all, since we do not yet include radiative acceleration and dust formation in our simulations, increasing the mass-loss rate only increases the density. If radiation pressure on dust grains is taken into account, the larger opacities expected when using higher mass-loss rates can alter the resulting morphology. Including more adequate descriptions of radiation shows that the discrepancy between the free wind approach and detailed descriptions increase with the mass-loss rate, and is small at lower mass-loss rates. This shows that the free wind approach is most viable for these lower mass-loss rates \citep{esseldeurs}. Secondly, the photodissociation radius strongly depends on the density, and in turn on the mass-loss rate. Increasing the mass-loss rate therefore also leads to larger models, and thus to a strong increase in the computation time of our models (see also Section \ref{section:pd}). 

In earlier studies of hydro-simulations of AGB stars , for example \citet{JM2}, particles reaching a critical radius were deleted from the simulation to increase computational efficiency. This allowed the authors to use higher resolutions, whilst keeping computation time moderate. Since we are interested in low-$J$ spectral lines, the size of the emitting envelope is of extreme importance, and we cannot use an outer boundary. In order to keep computational cost moderate, we therefore have to lower the resolution of our simulations. We have run test simulations with high resolution as well, and did not find significant differences in the inner wind morphology, as all the characteristic features of the high-resolution run are present in the low-resolution runs.

For all our models, the primary AGB star has an initial mass of $M_\mathrm{p}$ = 1.5 M$_\odot$, an effective radius of $R_\mathrm{p, eff} =$ 1.267 au, an accretion radius of $R_\mathrm{p, accr}$ = 1.220 au and a temperature of  $T_\mathrm{p, eff}$ = 2750 K. The secondary has a mass of $M_\mathrm{s}$ = 1 M$_\odot$ and an accretion radius of $R_\mathrm{p, accr}$ = 0.2 au. This accretion radius is rather large for the companion, but this was necessary to remove numerical instabilities as a result of the lower resolution. A lower value for the accretion radius of the secondary, in combination with a higher resolution, is especially needed when interested in resolving the accretion disc \citep[see][]{JM2}, at the cost of a higher computation time. Since we are looking at the simulations on a larger scale, we are not interested in resolving the accretion disc in great detail. This will not impact the resulting line profiles, as the region inside the accretion radius is negligible compared to the total region probed by the low-$J$ CO spectral lines. Additionally, if thermal dissociation were to be taken into account, we do not expect any CO to be present in the disc. The set-up parameters are summarised in Table~\ref{tab:parameters}.

We consider three different orbital separations of 9, 15, and 25 au, corresponding to orbital periods of approximately 17, 37, and 72 yrs, respectively. All models are assumed to be circular (i.e. $e = 0$). Lastly, we take three different initial velocities $v_\mathrm{ini}$ for the free wind, of 5, 10, and 20 km s$^{-1}$. The list of models are summarised in Table~\ref{tab:orbital}. It should be noted that for the higher outflow velocities, observations suggest a higher mass-loss rate \citep{2020A&A...640A.133R}. 
However, to explore the parameter space these models still remain interesting to study and understand.

We run the simulations for the v05 and v10 models until $\approx 2560$ yr, and the v20 models until $\approx 1910$ yr (see Section \ref{section:pd}). The number of particles in each simulation varies with the initial velocity, as the time between consecutive shells of injected particles can be written as \citep{siess}
\begin{equation}
    \delta t =  w_{ss} \frac{d_\perp}{v_\mathrm{ini}} 
\end{equation}
with $d_\perp$ the distance between neighbouring particles on an injection shell, $w_{ss}$ a value close to 1, and ${v_\mathrm{ini}}$ the initial wind velocity. This means that increasing the initial velocity of the wind, decreases the time between consecutive shell ejections, leading to more particles. At the end, the models with initial wind velocities of 5, 10, and 20 km s$^{-1}$ have approximately $2 \times 10^6$, $4 \times 10^6$, and $6 \times 10^6$ particles, respectively. 

\begin{table}[hbtp]
    \caption{\label{tab:parameters} Overview of the initial set-up parameters. }
    \begin{center}
        \begin{tabular}{l l l} 
        \hline
        \hline
        Parameter &  Unit  & Value \\
        \hline 
         $M_\mathrm{p}$ & [M$_\odot$] & 1.5 \\ 
         $T_\mathrm{p, eff}$ & [K] & 2750 \\
         $R_\mathrm{p, eff}$ & [au] & 1.267 \\
         $R_\mathrm{p, accr}$ & [au] & 1.220 \\
         $R_\mathrm{inject}$ & [au] & 4.259 \\
         $M_\mathrm{s}$ & [M$_\odot$] & 1 \\ 
         $R_\mathrm{s, accr}$ & [au] & 0.2 \\
         $\Dot{M}$ & [M$_\odot$ yr$^{-1}$] & $10^{-7}$ \\
         $t_\mathrm{max}$ & [yr] & 2560 / 1960 \\
        \hline 
    \end{tabular}
    \end{center}
    {\textbf{Notes.} \footnotesize{The set-up parameters are the same for all models: 
    $M_\mathrm{p}$, $M_\mathrm{s}$, $R_\mathrm{p, acc}$, and $R_\mathrm{s, acc}$ are the initial masses and accretion radii of the primary (AGB) and secondary (companion) star of the binary; $\dot{M}$ is the mass-loss rate of the AGB star; $T_{\rm p, eff}$ is its effective temperature; and $R_\mathrm{p, eff}$ is its radius. The last parameter $t_{\rm max}$ is the end time of the simulation.}}
\end{table}

\begin{table}[hbtp]
    \caption{\label{tab:orbital} Binary model characteristics. }
    \begin{center}
        \begin{tabular}{c c c} 
        \hline
        \hline
        Model name &  $v_\mathrm{ini}$ [km s$^{-1}$]  & a [au] \\
        \hline 
         v05a09 & 5 & 9 \\ 
         v05a15 & 5 & 15 \\
         v05a25 & 5 & 25 \\ 
         v10a09 & 10 & 9 \\
         v10a15 & 10 & 15 \\
         v10a25 & 10 & 25 \\ 
         v20a09 & 20 & 9 \\
         v20a15 & 20 & 15 \\
         v20a25 & 20 & 25 \\ 
        \hline 
    \end{tabular}
    \end{center}
    {\textbf{Notes.} \footnotesize{Binary models with their characteristic input values. The model names are set in such a way that the characteristics can be deduced from it, with ‘v…’ denoting the input wind velocity in km s$^{-1}$ and ‘a…’ denoting the orbital separation in au.}}
\end{table}

\subsection{Radiative transfer}

To post-process our hydrodynamic simulations, we use \textsc{Magritte}\footnote{ \url{https://github.com/Magritte-code/Magritte}, version 0.9.9} \citep{Magritte1, Magritte2, 2022JOSS....7.3905D, CEULEMANS2024100889}, an open-source 3D NLTE line radiative transfer code. \textsc{Magritte} solves the radiative transfer equation along a fixed set of rays using a second-order Feautrier scheme. The ray tracing algorithm is grid-agnostic, and can easily cope with SPH models as input. For the CO abundance, we assume a constant fractional number density of CO to hydrogen of $1.5 \times 10^{-4}$, i.e. 
\begin{equation}
    \frac{n_{\mathrm{CO}}}{n_{\mathrm{H}} + 2 n_{\mathrm{H}_2}} = 1.5 \times 10^{-4},
\end{equation}
such that when all hydrogen is in $\mathrm{H}_2$, we have a constant fractional abundance of $3 \times 10^{-4}$, which is a commonly assumed value for oxygen-rich AGB stars \citep{teysier}, but this value can be as high as $ 8 \times 10^{-4}$ for carbon-rich AGB stars \citep{CO_n}. The number densities of $\mathrm{H}$ and $\mathrm{H}_2$ are calculated using equilibrium chemistry, such that the following condition holds: 
\begin{equation}
    P_{\mathrm{H}_2} = P_{\mathrm{H}}^2 K_{\mathrm{H}_2}.
\end{equation}
Here $K_{\mathrm{H}_2}$ is the dissociation constant of the reaction given by 
\begin{equation}
    K_{\mathrm{H}_{2}} = \mathrm{exp} \left( \frac{-\Delta G_{\mathrm{H}_{2}}}{\mathcal{R} T_g} \right)
\end{equation}
with $\mathcal{R}$ the universal gas constant, and $\Delta G_{\mathrm{H}_{2}}$ the change in Gibbs energy.  It follows that 
\begin{equation}
    n_{\langle \mathrm{H} \rangle} k T_g = P_{\mathrm{H}} + 2P_{\mathrm{H}}^2 K_{\mathrm{H}_{2}}.
\end{equation}
Once the partial pressures are known, the number densities are found using $ n_i = \frac{P_i}{k T_g}$. After this, CO photodissociation is applied to obtain a realistic CO profile (see Sect.~\ref{section:pd}).
Additionally, we include a turbulent velocity field of 1~km~s$^{-1}$ throughout the domain in our \textsc{Magritte} simulations, which impacts the line broadening. This value is in line with previous studies, although it can be taken as high as 2 km s$^{-1}$ \citep[see e.g.][]{turb2, turb1, De_Beck_2012}.

In our NLTE calculations, we take into account the first 40 rotational transitions of CO in the ground vibrational state. All the molecular data are taken from the Leiden atomic and Molecular Database\footnote{\url{https://home.strw.leidenuniv.nl/~moldata/}} \citep{2005A&A...432..369S, Leiden_database}. The collisional rate coefficients are adopted from \citet{yang_2010}. For all v05 and v10 models, we add a spherical outer boundary inside \textsc{Magritte} at 5000 au, and for the v20 models this is located at 6000 au. This is done in order to remove the non-physical outer regions of the SPH model where the velocities are very high due to the expansion into vacuum. These boundaries are, however, well outside the emitting CO region (see Section \ref{section:pd} and Fig.~\ref{Fig:global_morph}), and thus do not impact the line profiles. We add the cosmic microwave background radiation at the outer boundary of the model, and add a spherical inner boundary on the surface of the AGB star (i.e. on $R_\mathrm{p, eff}$), which is emitting as a blackbody with a temperature of $T_\star$. We always put the observer at a distance of 500 pc, which is a realistic distance for the observed AGB stars \citep{dist}, changing it would uniformly scale the observed flux proportional to the ratio of the distances squared. Finally, we always use velocity bin sizes of 0.1 km s$^{-1}$ for the spectral lines. 

The region between the injection radius and the surface of the AGB star is left empty, but can still contribute to the spectral lines, albeit towards the higher $J$ lines. We sample this region before calculating the spectral lines. Since many uncertainties are present regarding the complex structure, and a detailed treatment is out of the scope of this paper, we use simple prescriptions for the quantities in this region, similar to \citet{2006}. As stated in Section~\ref{section:gridsetup}, we assume a power law for the temperature according to Eq.~\ref{eq:Tr}. The density is given by the mass conservation equation
\begin{equation}
\label{eq:rho}
    \rho(r) = \frac{\dot{M}}{4\pi r^2 v(r)}, \ R_\star < r < R_\mathrm{inject}
\end{equation}
and the velocity is assumed to follow a classical $\beta$ law of the form 
\begin{equation}
    v(r) = v_\infty  \left( 1 - \frac{R_\star}{r}  \right)^\beta,  \ R_\star < r < R_\mathrm{inject} 
\end{equation}
with $\beta = 0.5$, and $v_\infty$ such that $v(r)$ equals the initial wind velocity of the model at $r=R_{\mathrm{inject}}$. In the inner region, the CO number density is calculated as explained above.

Because of the computational cost that comes with full NLTE calculations, \textsc{Magritte} can recursively remesh the grid, thereby reducing the computational load \citep[see][for details]{CEULEMANS2024100889}. This method has two parameters, the maximum variation of the density in a box $r_{\mathrm{max}}$, and maximum recursion level $l$. The exact number of points that are left after reduction depends on these two parameters, as well as the original model. We opt for a value of $l = 15$, and increase the $r_{\mathrm{max}}$ parameter until we are left with one quarter of the original number of points. This was found to be optimal for reducing the computational cost without sacrificing accuracy. Additionally, we benchmarked our results to the Monte-Carlo radiative transfer code \textsc{SKIRT} \citep{skirt}, for which the results can be found in Appendix~\ref{appendix D}.

\subsection{CO Photodissociation}
\label{section:pd}

When modelling low-$J$ CO transitions, the size of the envelope is of high importance, as it greatly influences the observed flux coming from the lowest transitions. This size is set by external ultraviolet (UV) radiation, that destroys CO from outside inwards, determining the CO distribution. When modelling CO spectral lines using spherically symmetric models, one can use the tabulated values of CO density profiles, as calculated by e.g. \cite{Mamom}, and more recently \citet{Groenewegen_2017} and \citet{saberi}. 

When dealing with more complex morphologies, the spherically symmetric description is insufficient, since the shielding from UV radiation by CO itself, and by H$_2$, varies depending on the direction of the incoming UV photons due to the flattening of the model and local density variations in a spiral structure. Therefore, to calculate the CO density profiles, we implement photodissociation in 3D, which closely follows the method described by \citet{Groenewegen_2017}, which is recapitulated and generalised below.

Assuming no molecules are created, and UV photons are the only destruction mechanism, the CO number density, $n$, along a ray can be written as \citep{1981ApJ...251..181J}
\begin{equation}
    \frac{1}{r^2}\frac{\partial}{\partial r} \left(r^2 n v(r) \right) = -I(r)n,
\end{equation}
where $v(r)$ is the expansion velocity, and $I(r)$ is the photodissociation rate at a distance $r$ to the central AGB star. In order to find a solution, we define $x$ as $r^2 n v(r)$, leading to
\begin{equation}
\label{eq:sol}
    x(r_i) = x(r_{i-1}) \ \mathrm{exp} \left( - \int_{r_{i-1}}^{r_i} \frac{I(r')}{v(r')} \mathrm{d}r' \right),
\end{equation}
where one assumes $x(r_0) = 1$. The dissociation rate $I(r)$ is given by 
\begin{equation}
\label{eq:Ir}
    I(r) = \frac{1}{4 \pi} \int k(r, \theta, \phi) \mathrm{d}\Omega,
\end{equation}
where $k(r, \theta, \phi)$ is the dissociation rate at a distance $r$ from interstellar photons that are coming at a direction defined by $\theta$ and $\phi$ towards point $r$. In the spherically symmetric case, the integral only needs to run over $\theta$. The dissociation rate  $k(r, \theta, \phi)$  can be written as 
\begin{equation}
\label{eq:kr}
     k( r, \theta, \phi) = \chi \ I_0 \ \Theta_{\mathrm{dust}} ( r, \theta, \phi) \ \Theta_{\mathrm{H}_2, \mathrm{CO}} ( r, \theta, \phi),
\end{equation}
where $I_0 = 2.6 \times 10^{-10}$ s$^{-1}$ is the unshielded dissociation rate \citep{visser}, $\chi$ indicates the strength of the interstellar radiation field (ISRF) relative to the Draine field \citep{Draine} used in \citet{visser}, $\Theta_{\mathrm{dust}}(r, \theta, \phi)$ represents the amount of shielding by dust, and $\Theta_{\mathrm{H}_2, \mathrm{CO}}(r, \theta, \phi)$ corresponds to the CO self shielding and H$_2$ shielding. For the latter, we use the tabulated shielding functions of \citet{visser}, which depend on the CO and H$_2$ column densities, and the excitation temperature $T_\mathrm{ex}$, and are interpolated accordingly. We assume the excitation temperature equals the kinetic temperature. The shielding by dust is given by $\Theta_{\mathrm{dust}} = \mathrm{exp} (- \tau_{\mathrm{dust}})$, with $ \tau_{\mathrm{dust}}$ the dust optical depth. In our \textsc{Phantom} models, we do not yet include dust; therefore, we opted to consider a constant dust extinction at 1000 \AA, given by 
\begin{equation}
    \tau_{\mathrm{dust}}( r, 1000 \text{\AA}) = \frac{4.65 \times 2 \times N_{\mathrm{H}_2}(r)}{1.87 \times 10^{21}}
\end{equation}
with $N_{\mathrm{H}_2}(r)$ the H$_2$ column density \citep{1983ApJ...264..546M}. Finally, we assume $\chi =1$.

Since the dissociation rate depends on the CO column density, which in turn depends on the CO density profile, we need to solve this iteratively. Since we now work in 3D, this would entail tracing a ray through every point, and for each point along that ray, calculate the column densities along all directions, and finally solve Eq.~\ref{eq:sol} along all the rays. As this is computationally demanding, we instead opt for an approximation, where we apply the process described above for a predetermined amount of rays, which are subsequently interpolated to obtain values of $x(r)$ for the full domain, similar to \citet{esseldeurs}. The full implementation is outlined in Appendix~\ref{appendix A}. 

One also has to keep in mind the initial size of the model, as the H$_2$ dissociation is not being accounted for, and thus, the H$_2$ and dust shielding is always increasing. In \citet{Groenewegen_2017}, the size of the model is taken to be $15 R_{\frac{1}{2}}$ (i.e. 15 times the radius where the fractional abundance of CO is half its central value), but running our hydro-simulations to such a size is computationally unfeasible. Instead, we run the model sufficiently large, such that the value of $R_{\frac{1}{2}}$ does not change by more than one percent. We find that running the v05 and v10 models to 2560 years suffices for this, whilst for the v20 models, this state is reached sooner, and we stop the simulations at 1910 years. Running the model even further could still lead to slightly different profiles in the outer regions, but this should only marginally impact the $J = 1{\rightarrow} 0$ line, and this is prone to larger uncertainties such as the exact value of $\chi$. 

\section{Results}
\label{section:results}
\begin{figure*}[ht]
\centering
	\includegraphics[width = 0.965\textwidth]{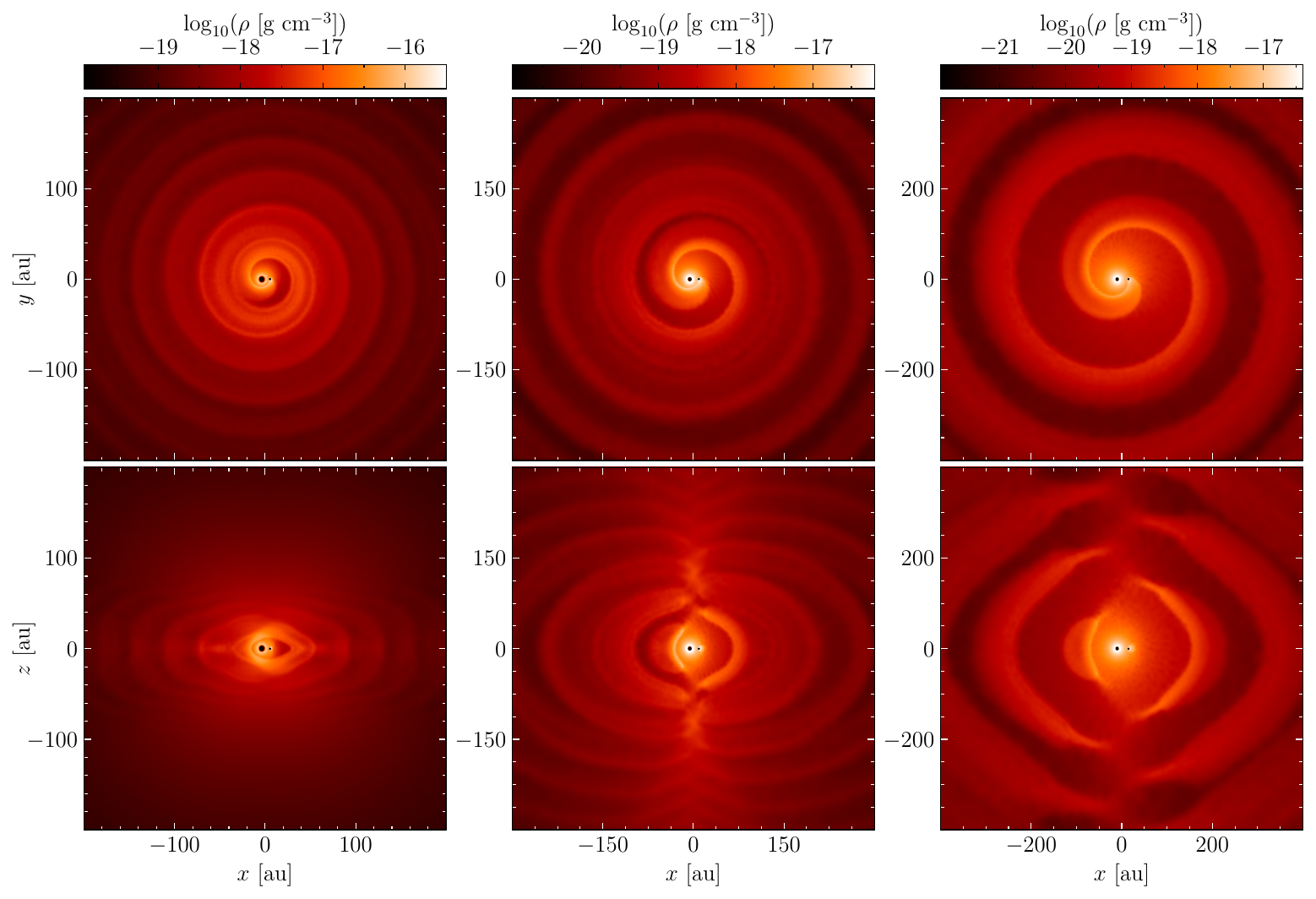}
	\caption{ Density distributions in slices through the orbital plane (upper row) and the meridional plane (lower row) for the models with initial wind velocities of 10 km s$^{-1}$. From left to right, the orbital separation is 9, 15, and 25 au. }
	\label{Fig:morphologies_v10}
\end{figure*}
\begin{figure*}[ht]
\centering
	\includegraphics[width = 0.965\textwidth]{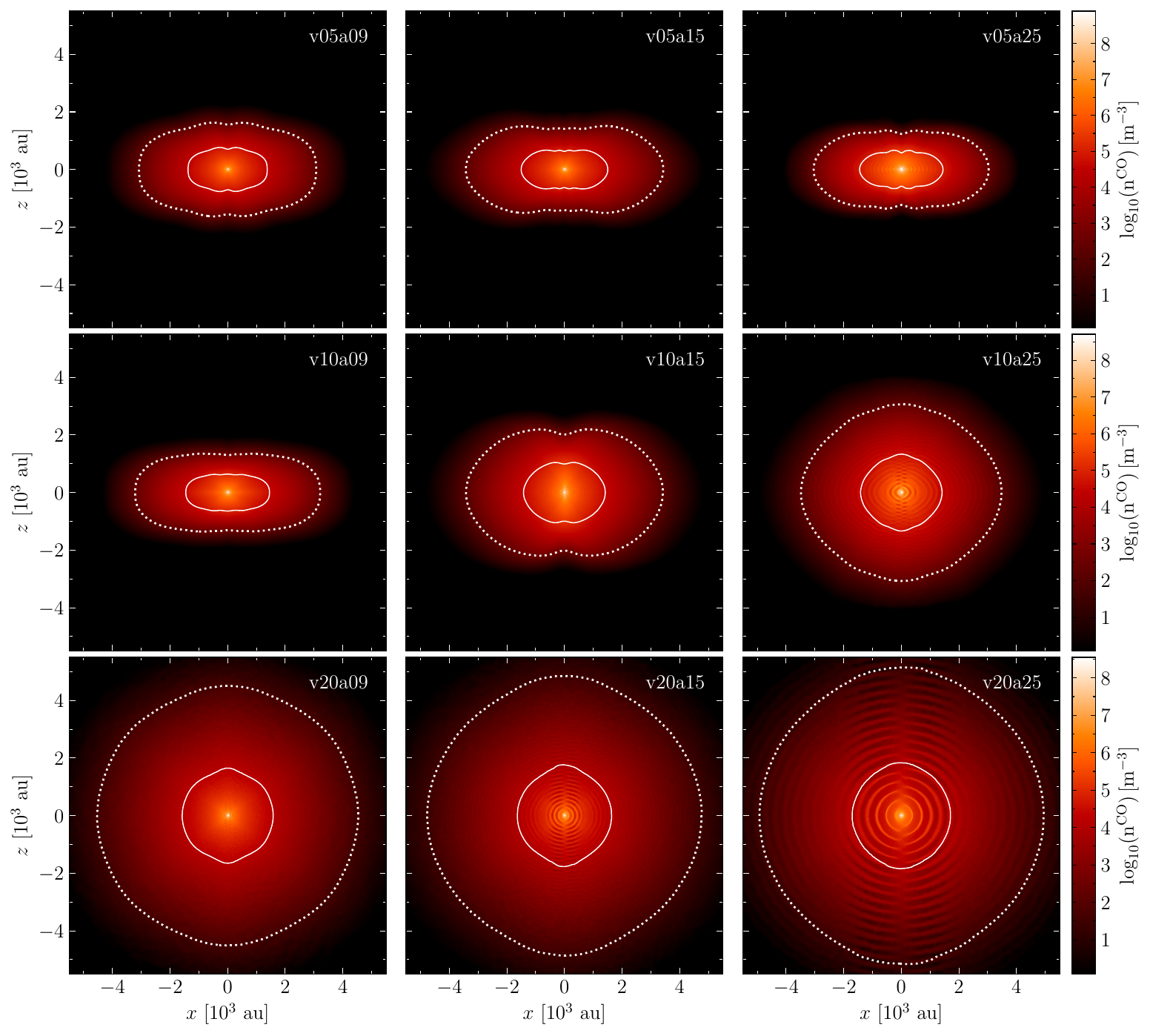}
	\caption{Number density of CO in slices through the meridional plane after accounting for photodissociation. The solid and dotted lines represent the location where $x(r)$ is 0.5 and 0.01, respectively.}
	\label{Fig:global_morph}
\end{figure*}

\subsection{Morphologies}
\label{section:morph}

As a novel part of the wind parameter space is explored, the different resulting morphologies are briefly explained. Still, the general structures that are observed are similar to those found in earlier studies. Thus, we omit a detailed description of the formation of these structures, and for a far more in-depth explanation of how these specific morphologies are created, we refer to \citet{SM1} and \citet{JM1, JM2} and references therein. We focus here on the inner region of the models first, and discuss the overall global morphology (after we have applied CO photodissociation) in the subsequent section.

\subsubsection{Inner wind morphology}

A companion influences the outflow in two main ways. First, it interacts with the outflowing wind particles due to gravitational attraction. Secondly, it induces an orbital motion of the AGB star around the CoM. We show the resulting mass density for the v10 models in slices through both the orbital ($x-y$) and meridional ($x-z$) plane for the inner regions in Fig.~\ref{Fig:morphologies_v10}, with increasing orbital separation from left to right. Analogous plots for the v05 and v20 models can be found in Fig.~\ref{Fig:morphologies_v05} and Fig.~\ref{Fig:morphologies_v20}. 

For the initial velocity of 10 km s$^{-1}$ (Fig.~\ref{Fig:morphologies_v10}) a clear spiral structures arises in the orbital plane, regardless of the separation. In all three cases, the spiral consists of a more dense inner edge, the `backward spiral edge' (BSE), and a less dense outer edge, the `frontal spiral edge' (FSE) as defined in \citet{JM1}. This FSE has a greater radial velocity, and therefore the spiral wake will widen, eventually catching up with the BSE of the spiral sent out earlier, and form a single spiral shock. This interaction is more pronounced at the smaller separations.

Although mostly similar in the orbital plane, the morphologies in the meridional plane are very distinct. Looking at the largest separation of 25 au (right column), the density distribution consists of concentric arcs, where the inner part of each arc is the BSE, and the outer edge is the FSE. At the ends of the arcs, there are thinner regions, resulting from particles that are not gravitationally attracted by the companion. At a separation of 15 au (middle column), we still see the same concentric shells, but the catching up of the FSE and BSE occur sooner, removing also the thin ends of the arcs. At the smallest separation of 9 au (left column), the arcs have a very distinct shape, being elongated in the orbital plane, indicating the spiral does not reach very high above the orbital plane.  

Compared to the lower initial outflow velocity of 5 km s$^{-1}$ (Fig. \ref{Fig:morphologies_v05}), the spiral structures look very similar to the v10a09 model, where the FSE quickly catches up with the BSE. One notable difference is the appearance of a bow-shock spiral for the v05a15 and v05a25 models. This can be seen by the presence of a second spiral structure arising in front of the companion in the outer region of the regular spiral. Interesting to note is the absence of this feature in the v05a09 model, as it is expected to be more present when the wind velocity at the location of the companion is lowest. In the meridional plane, we see structures that are reminiscent of those observed for the v10a09 model, that are elongated along the orbital axis, although the exact shape differs between the models.

For the highest initial outflow velocity of 20 km s$^{-1}$ (Fig.~\ref{Fig:morphologies_v20}) we see a very well-defined narrow spiral structure in the orbital plane for all three models. Still there is a FSE and BSE present; however, only at the smallest separation can we observe the widening of the spiral and the FSE catching up with the BSE at this scale. In the meridional plane, the concentric shells reach nearly up to the polar axis, and are not too dissimilar from those observed in the v10a25 model, although the shells appear more circular, whilst those in the v10a25 model are more triangular, because of a stronger interaction of the companion with the wind.

\subsubsection{Large scale morphology and CO photodissociation}
\begin{figure*}[ht]
\centering
	\includegraphics[width = 0.965\textwidth]{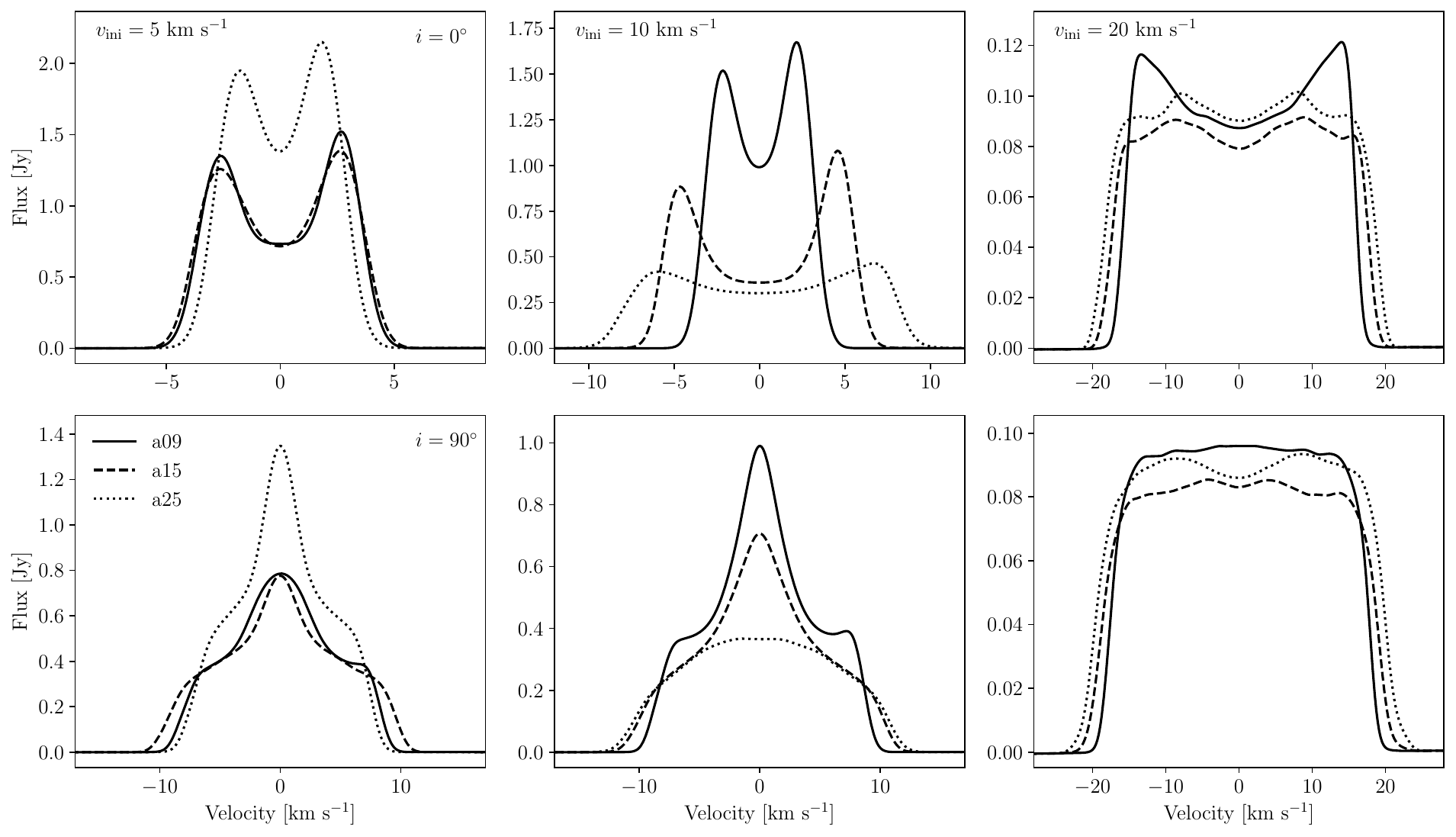}
	\caption{Synthetic CO $J = 2{\rightarrow}1$ spectral lines for all models. The columns correspond, from left to right, to initial outflow velocities of 5, 10, and 20 km s$^{-1}$.  The solid, dashed, and dotted lines represent orbital separations of 9, 15, and 25 au. The spectral lines are creating viewing face-on at an inclination of $0 \degree$ (upper row) and edge-on at $90 \degree$ (lower row).}
	\label{Fig:CO2-1_i=0}
\end{figure*}

The global morphologies of all nine models, after the calculation of the CO number density, are displayed in Fig.~\ref{Fig:global_morph}, where the solid and dotted lines represent the location where fractional abundance of CO with respect to H$_2$, compared to the initial fractional abundance, is 0.5 and 0.01, respectively. We always show a slice through the $x-z$ plane, as in the orbital plane the CO map is circular in all models. In the meridional plane, some  interesting features are visible. Due to the motion of the AGB star about the CoM, the wind particles are injected into the simulation with a velocity
\begin{equation}
    \boldsymbol{v}_\mathrm{wind} = \boldsymbol{v}_\mathrm{ini} + \boldsymbol{v}_\mathrm{orb, AGB}
\end{equation}
with $v_\mathrm{ini}$ the injection velocity, and $ v_\mathrm{orb, AGB}$ the orbital velocity of the AGB star around to CoM. As noted in \citet{JM1}, the wind particles ejected perpendicular to the orbital plane follow a radial direction with an angle
\begin{equation}
    \theta = \tan^{-1}\left( \frac{\boldsymbol{v}_\mathrm{ini} }{\boldsymbol{v}_\mathrm{orb, AGB}} \right)
\end{equation}
with respect to the orbital plane. This deflection of the particles' trajectory causes the global morphology to be flattened. This is readily observed in Fig.~\ref{Fig:global_morph}, where the v05 models all show very flattened morphologies, whilst the v20 models are more spherical. For the v10 models, flattening is strong for the v10a09 model, only marginal for the v10a15 model, and very small for the v10a25 model, which is in line with expectations as the orbital velocity around the CoM is lowered for larger separations. In the meridional plane, the arcs also reach lower latitudes and are elongated along the orbital axis with decreasing separation. The gravitational attraction of the companion can further focus particles to the orbital plane, building up pressure gradients that can cause particles to be accelerated towards lower density regions. At larger scales, we also see that the FSE still catches up with the BSE for the larger separations for the v20 models.

Due to UV photodissociation, the fractional CO abundance in the orbital plane, with respect to the central fractional abundance, drops to half for all models between 1500 and 2000 au, and for the flattened morphologies, this is already at $\sim 1000$ au in the $z$ direction. This inner part is important for most CO transitions as they are only excited within this region, and the outer regions only impact the $J = 1{\rightarrow}0$ and marginally impact the $J =2{\rightarrow}1$ lines. The CO fractional abundance, relative to the central value, drops to 1 percent at $\sim 4000$ au for the v05 and v10 models in the orbital plane, and 2000 au in the $z$ direction. For the v20 models this occurs around 5000 au. 

Overall, the global CO distribution closely follows the inner wind morphology, as the shape of the $R_{\frac{1}{2}}$ contour (or where the fractional abundance of CO is half of its initial value) closely resembles the shape of the arcs in the meridional plane, which is best visible for the v05 models. Even for the v20 models, where no flattening occurs, the contours show deviations from spherical symmetry with small bumps towards the polar regions. This also shows the importance of accounting for the CO photodissociation in a detailed manner. Finally, we note that when looking at the global morphology, the interaction of the binary companion with the wind (besides the global flattening), is hardly visible for the smallest separations, and becomes more visible with increased separations.

\subsection{Synthetic spectral lines}
\label{section:synthetic spectral lines}

For the CO spectral lines, not only the density structures, but also the velocity distribution is important, as the latter dictates the frequency at which certain material will be emitting. Additionally, the temperature profile is of great relevance, since this greatly impacts the level populations. Given the asymmetry of our simulations, we create spectral lines at six different inclinations between $0 \degree$ and $90 \degree$, where $0 \degree$ is defined as looking straight at the orbital plane, or face-on, whilst for $90 \degree$ we are looking at the meridional plane, or edge-on. When viewing at an angle, the position angle (PA) of the mass-losing AGB star also can play a role, as it changes the appearance of the spiral structure. We define a PA of $0 \degree$ when both the AGB star and the companion are on the $x$-axis, with the AGB star on the left (i.e., as they are presented in Fig.~\ref{Fig:morphologies_v10}), and anti-clockwise is positive. Our focus here lies on the low-$J$ lines, starting from the $J = 1{\rightarrow}0$ up to the $J = 6{\rightarrow}5$ transition, as these lines are often used to derive mass-loss rates from observations. In the following sections, any telescope effects are omitted. These will be discussed in Sect.~\ref{section:beam} and Sect.\ref{section:noise}.

\begin{figure*}[ht]
\centering
	\includegraphics[width =0.965\textwidth]{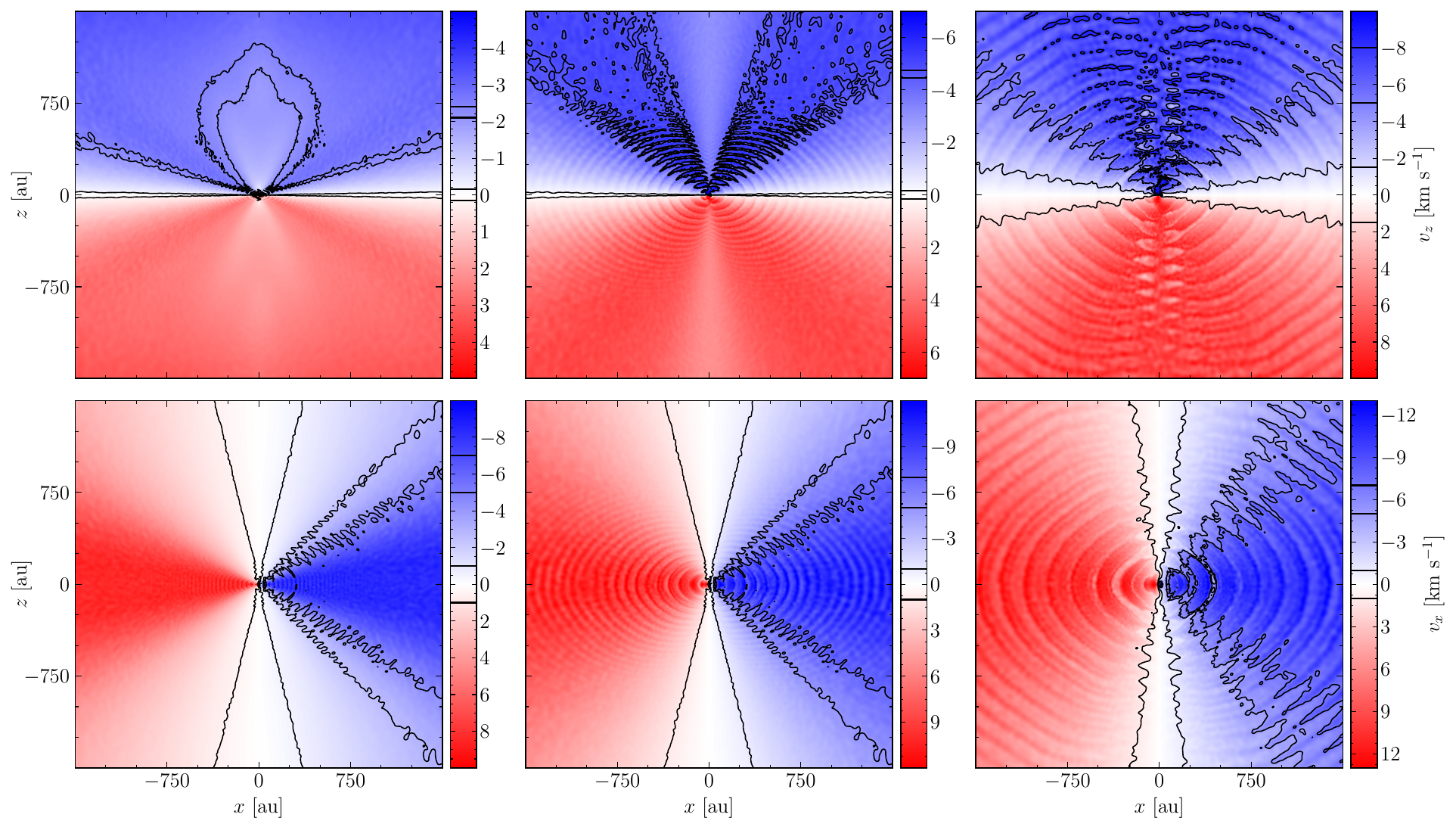}
	\caption{ Velocities in the $z$ direction (upper row) and in the $x$ direction (lower row) in slices through the meridional plane for the models with initial wind velocities of 10 km s$^{-1}$. Contours are shown on the plots around velocities interesting for the line profiles, indicated by the horizontal lines on the colour bar. From left to right, the orbital separation is 9, 15, and 25 au.}
	\label{Fig:velocities_v10}
\end{figure*}

\subsubsection{Face-on and edge-on line profiles}
\label{section:lineshapes}
\begin{center}
    \begin{figure}[ht]
    \centering
	\includegraphics[width = 0.9\columnwidth]{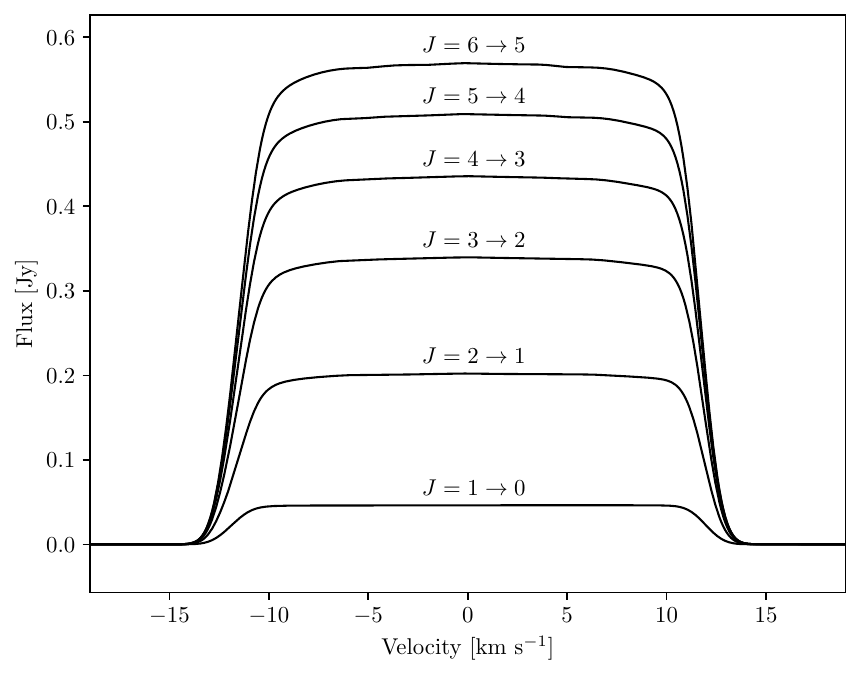}
	\caption{Synthetic CO spectral lines of the first six transitions from a single-star \textsc{Phantom} model, with a mass-loss rate of $10^{-7}$ M$_\odot$ yr$^{-1}$ and an initial wind velocity of 10 km s$^{-1}$. }
	\label{Fig:1D_lines}
    \end{figure}
\end{center}

Fig.~\ref{Fig:CO2-1_i=0} shows the CO $J = 2{\rightarrow}1$ spectral lines for all nine models viewed both face and edge-on, always at a PA of $0 \degree$. As noted before, this line is excited throughout the wind, and thus should trace the overall morphology. For the v05 models (left column) very defined double peaks emerge at the maximal radial velocities when viewed face-on, and a strong central bump appears when viewed edge-on, regardless of the orbital separation.  These characteristics are are still present for the v10a09 and v10a15 models (middle column), with the central bump for the v10a09 model being very prominent. The two peaks still show up for the v10a25 model when viewing face-on, but the central bump has completely disappeared for this model, and the line profile approaches a simpler parabolic shape. Looking at the v20 models (right column), the profiles show less pronounced features. Some peaks are still present when viewed face-on, where, except for the v20a09 model, the peaks are no longer present at the outer region of the spectral line, but have moved closer to the central velocities. Viewing edge-on, the line profiles are approaching flat topped line profiles, characteristic of optically thin spherically symmetric outflows (see also Fig.~\ref{Fig:1D_lines}).

Additionally, we see an increase in the width of the spectral lines, when going from the face-on to the edge-on view. For the v05 models, the line width is increased by approximately a factor of two for the three orbital separations. For the v10 models, it changes between the models, with again a factor of two for the v10a09 model, 1.5 for the v10a15 model, and for the v10a25 the width has remained similar between the two inclinations. 

When comparing the different spectral lines more closely, we see that the flux changes significantly between the different models, especially when comparing different outflow velocities. This can be attributed to several reasons. First of all, increasing the wind velocity naturally widens the line profiles, but also decreases the density given a constant mass-loss rate (Eq.~\ref{eq:rho}), thereby lowering the amount of flux received per velocity bin. Secondly, the interaction with the companion also drastically alters the temperature of the wind, which in turn greatly affects the level populations, and therefore the received flux. As described in Section \ref{section:morph}, this modification strongly depends on both the local wind velocity as well as the orbital separation. 

To gain insight into the shapes of these spectral lines, we show the velocity distribution of the hydro simulations in the $z$ and $x$ direction (i.e., $v_z$ and $v_x$) in slices through the meridional plane. For the v10 models, we show this in Fig.~\ref{Fig:velocities_v10}. Analogous plots for v05 and v20 are presented in Fig.~\ref{Fig:velocities_v05} and \ref{Fig:velocities_v20}, respectively. In these plots, the observer is located above the plot in the upper row, and to the right for the lower row.  Additionally, two sets of contours are indicated to visualise velocity bins corresponding to interesting features of the spectral lines. For reference, in a spherically symmetric wind with a constant outflow velocity, the volume of equally sized velocity bins is equal and are conically shaped, leading to flat topped and parabolic line profiles in optically thin and thick regimes, respectively. Spectral lines of the first six transitions from a spherically symmetric \textsc{Phantom} model, with a mass-loss rate of $10^{-7}$ M$_\odot$ yr$^{-1}$ and initial wind velocity of 10 km s$^{-1}$ are shown in Fig.~\ref{Fig:1D_lines}. Note that it is also possible to obtain two peaked profiles from single star systems, depending on whether the outflow is resolved or not (see also Sect.~\ref{section:beam}).

Instead of the conically shaped velocity bins, the interaction with the companion for the v10a09 model causes material in the velocity bin around 2.2 km s$^{-1}$ to deviate strongly from the conical shape (as visible in the upper left plot of Fig.~\ref{Fig:velocities_v10}). In this case the velocity bin has an arc-like shape. The velocity bin centred around 0 km s$^{-1}$ does remain conical, and thus the relative contribution of the velocity bin centred around 2.2 km s$^{-1}$ will be greater, leading to an increase in the flux at $\pm 2.2$ km s$^{-1}$, and two peaks in the spectral line (Fig.~\ref{Fig:CO2-1_i=0} top middle plot, solid line). As seen in Fig.~\ref{Fig:velocities_v05}, this also produces the two peaks for all the v05 models, although the exact shape of the velocity bin structures strongly differs between the orbital separations. All the double peaks in Fig.~\ref{Fig:CO2-1_i=0} also show a stronger peak at redshifted velocities, in contrast with the blue shifted peak, which is due to the opacity being generated in slightly hotter regions for the red peak \citep{morris}. As we are looking face-on, this cannot be an effect of the position angle.

The central bump when viewing edge-on has a similar explanation as the appearance of two peaks. Although in this case all velocity bins are roughly conical, the volume of the bins towards the central velocity is larger compared to that of higher velocities. This is more outspoken for the v05 models (see Fig.~\ref{Fig:velocities_v05}). For these models, the velocity bins deviate more from the conical shape as well, as the contours show more curved edges, due to a stronger interaction with the companion. Note also the different scales of the colour bars in the top and bottom row. In the orbital plane, higher speeds are reached compared to the meridional plane, giving rise to the increase in the width of the spectral lines described above.

The two peaks that are present in the v10a15 model are not due to an arc-like structure, but rather due to two separate regions having the same projected velocity in the $z$ direction, thereby creating a peak in the line profile. The double peaks from the v10a25 model are yet again different in origin, this time they are created akin to the central bump feature. The conical regions are increased in size with respect to the central velocity bin (top right plot of Fig.~\ref{Fig:velocities_v10}), creating two distinct peaks. We also note that the individual contours show some coves, indicating that the features of the spiral are influencing the line shapes stronger than for the v05 models, where the contours are more smooth (see Fig.~\ref{Fig:velocities_v05}).

Finally, the cause of the deviations from flat topped or parabolic line profiles for the v20 models are similar in origin, but as these deviations are less pronounced, they are also less visible in Fig.~\ref{Fig:velocities_v20}. For the v20 models, however, the spiral features in the contours are very pronounced, which can impact the line profiles (see also Sect. \ref{section:higher co} and \ref{section:effect of PA}).

\subsubsection{Intermediate inclinations}
\label{section:intermediate inclinations}

\begin{figure}[t]
	\includegraphics[width = \columnwidth]{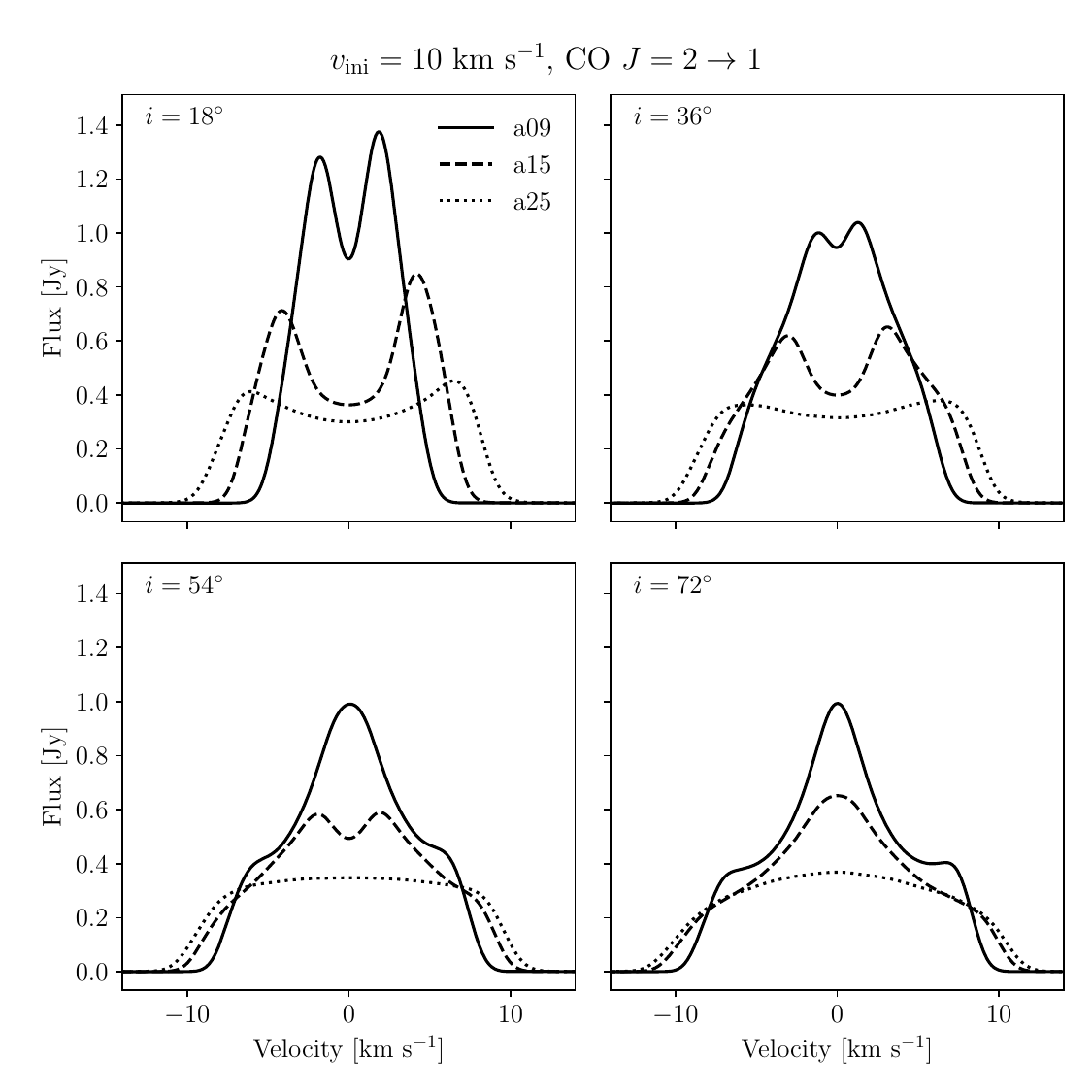}
	\caption{Synthetic CO $J = 2{\rightarrow}1$ spectral lines for the models with initial wind velocities of 10 km s$^{-1}$, viewed under inclinations of $18\degree$, $36\degree$, $54\degree$, and $72\degree$. The solid, dashed, and dotted lines represent orbital separations of 9, 15, and 25 au. }
	\label{Fig:inclinations_v10}
\end{figure}

In Fig.~\ref{Fig:inclinations_v10} we show the CO $J = 2{\rightarrow}1$ spectral lines for all v10 models viewed at intermediate inclinations of $18\degree$, $36\degree$, $54\degree$, and $72\degree$, still at a PA of $0 \degree$. Analogous plots for the v05 and v20 models are presented in Figs.~\ref{Fig:inclinations_v05} and \ref{Fig:inclinations_v20}, respectively. In general, there is a smooth transition between the face-on and edge-on cases, as at ever-increasing inclinations, the two peaks become less pronounced, and the central bump starts appearing. At the two most central inclinations of $36\degree$ and $54\degree$, both features can be distinguished. This is most visible at $36\degree$ for the v10a09 model and at $54\degree$ for the v10a15 model, where two small peaks are present on a larger central emission. The spectral lines at the smallest and largest inclinations of $18\degree$ and $72\degree$ are virtually indistinguishable from the face and edge-on spectral lines presented in Fig.~\ref{Fig:CO2-1_i=0}. The same findings apply to the v05 and v20 models.

\subsubsection{Different CO transitions}
\label{section:higher co}
\begin{figure}[t]
	\includegraphics[width = \columnwidth]{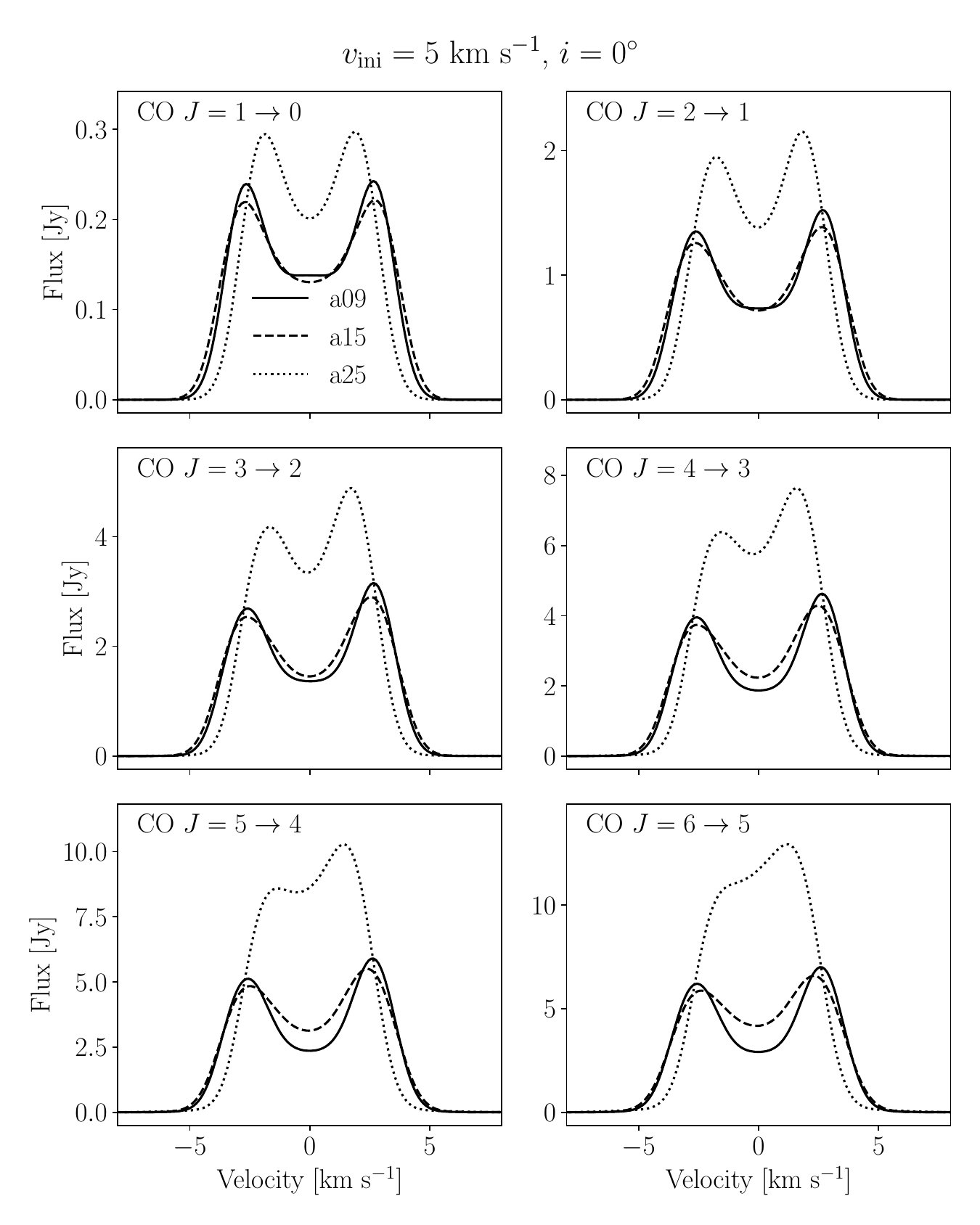}
	\caption{Synthetic CO spectral lines of the first six rotational transitions, for the models with initial outflow velocities of 5 km s$^{-1}$, viewed under an inclination of $0\degree$. The solid, dashed, and dotted lines represent orbital separations of 9, 15, and 25 au. }
	\label{Fig:trans_v5_i0}
\end{figure}

\begin{figure}[t]
	\includegraphics[width = \columnwidth]{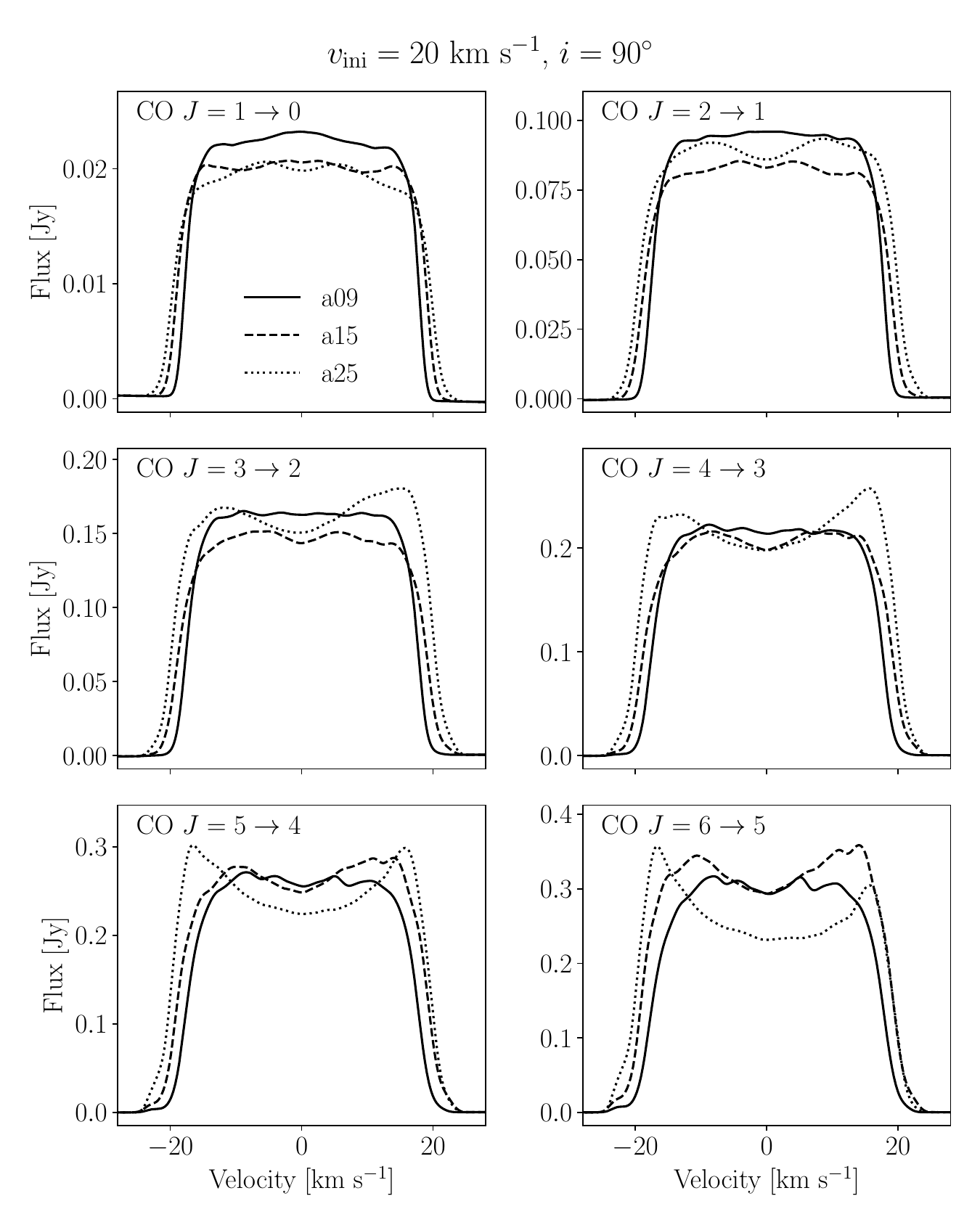}
	\caption{Synthetic CO spectral lines of the first six rotational transitions, for the models with initial outflow velocities of 20 km s$^{-1}$, viewed under an inclination of $90\degree$. The solid, dashed, and dotted lines represent orbital separations of 9, 15, and 25 au. }
	\label{Fig:trans_v20_i90}
\end{figure}

\begin{figure}[ht]
	\includegraphics[width = \columnwidth]{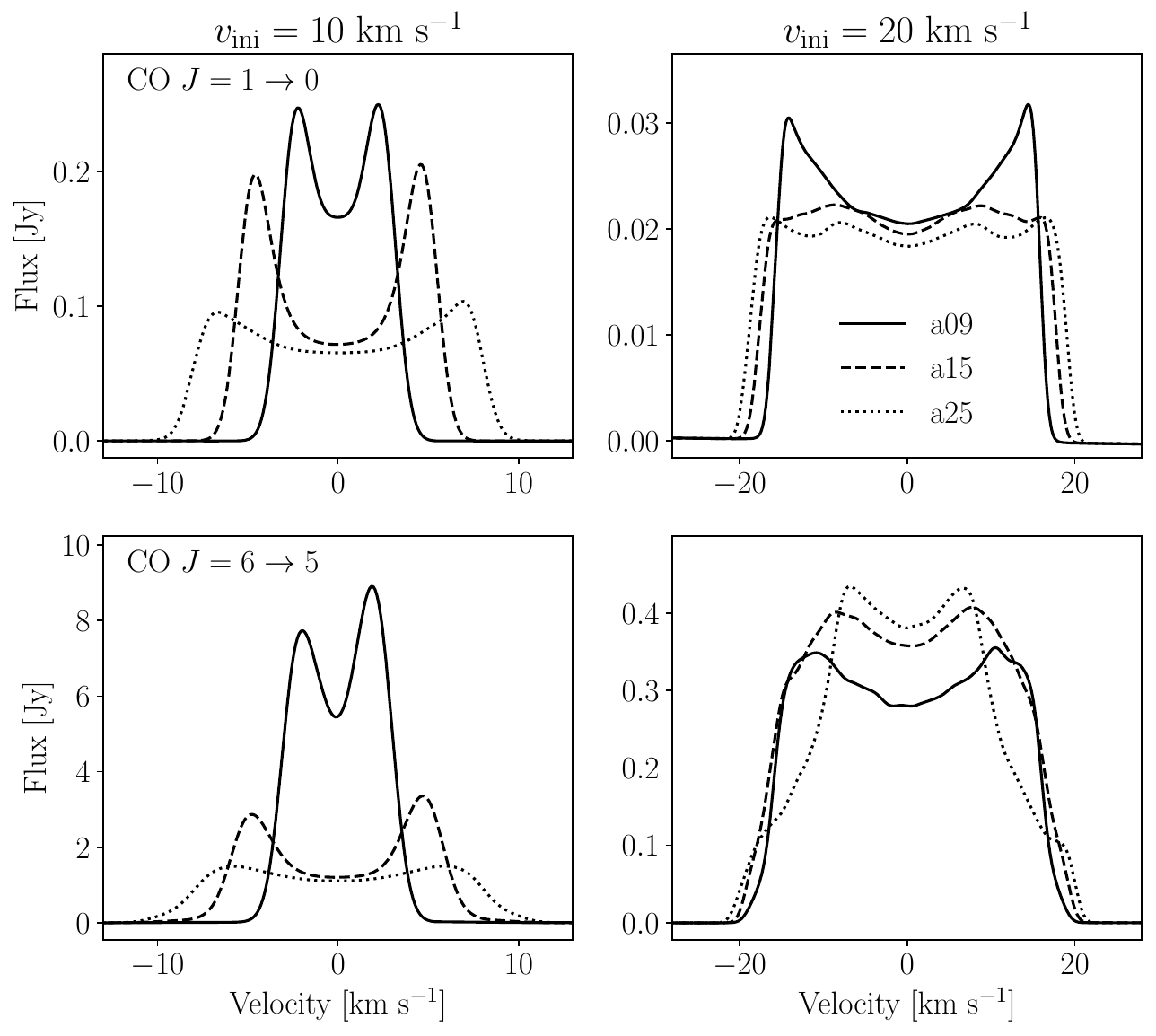}
	\caption{Synthetic CO $J = 1{\rightarrow}0$ (top) and $J = 6{\rightarrow}5$ (bottom) spectral lines for the v10 (left) and v20 (right) models, viewed under an inclination of $0\degree$. The solid, dashed, and dotted lines represent orbital separations of 9, 15, and 25 au. }
	\label{Fig:trans_i0}
\end{figure}

\begin{figure}[ht]
	\includegraphics[width = \columnwidth]{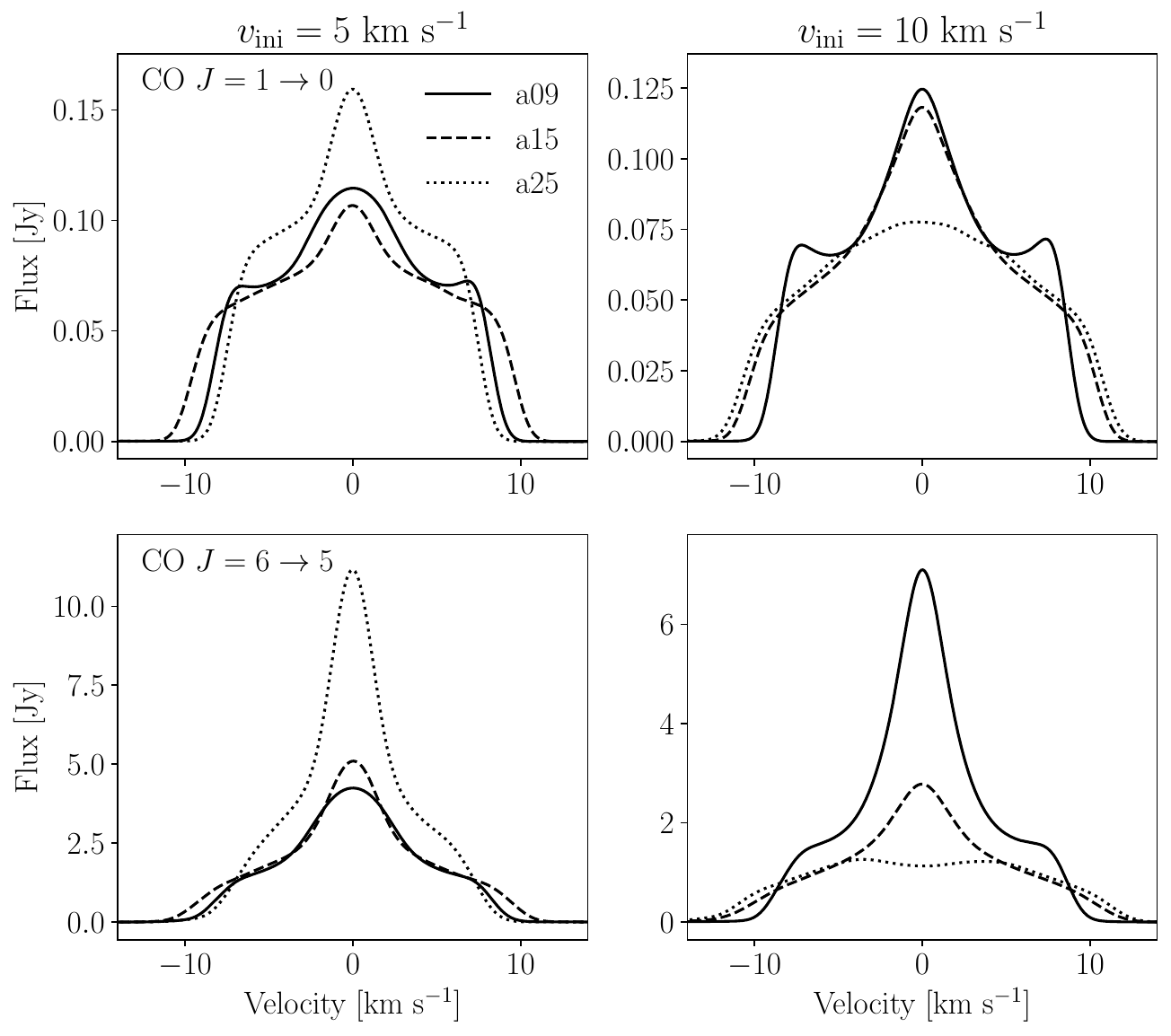}
	\caption{Synthetic CO $J = 1{\rightarrow}0$ (top) and $J = 6{\rightarrow}5$ (bottom) spectral lines for the v05 (left) and v10 (right) models viewed under an inclination of $90\degree$. The solid, dashed, and dotted lines represent orbital separations of 9, 15, and 25 au.  }
	\label{Fig:trans_i90}
\end{figure}

The CO $J = 2{\rightarrow}1$ line, which we have so far solely focussed on, probes mainly the outer part of the CSE. Higher transitions probe hotter regions due to a higher excitation temperature. In Fig.~\ref{Fig:trans_v5_i0} we show how the spectral lines evolve as the transition increases from $J=1 {\rightarrow}0$ up to $J = 6{\rightarrow}5$ for the v05 models viewed face-on, and in Fig.~\ref{Fig:trans_v20_i90} for the v20 models viewed edge-on. For the remaining models, we show only the $J = 1{\rightarrow}0$ and the $J = 6{\rightarrow}5$ transitions viewed both face-on and edge-on in Fig.~\ref{Fig:trans_i0} and Fig.~\ref{Fig:trans_i90}, as the results are similar to the cases described here.


Starting with the v05 models viewed face-on in Fig.~\ref{Fig:trans_v5_i0}, we see that the peaks remain strongly visible for the models with orbital separations of 9 and 15 au, although the central dip diminishes in strength for the 15 au case towards higher transitions. For the orbital separation of 25 au, at the highest considered transition the two peaks are replaced by a single off-centre peak, as the dip has disappeared completely. Remarkable is also that at the $J = 1{\rightarrow}0$ transition, the strength of both peaks is equal, due to a lower optical depth, and the effect described in Sect.~\ref{section:lineshapes} does not take place. In general, for the v05 models, the line shape remains largely similar when viewing face-on.

\begin{figure}[t]
	\includegraphics[width = \columnwidth]{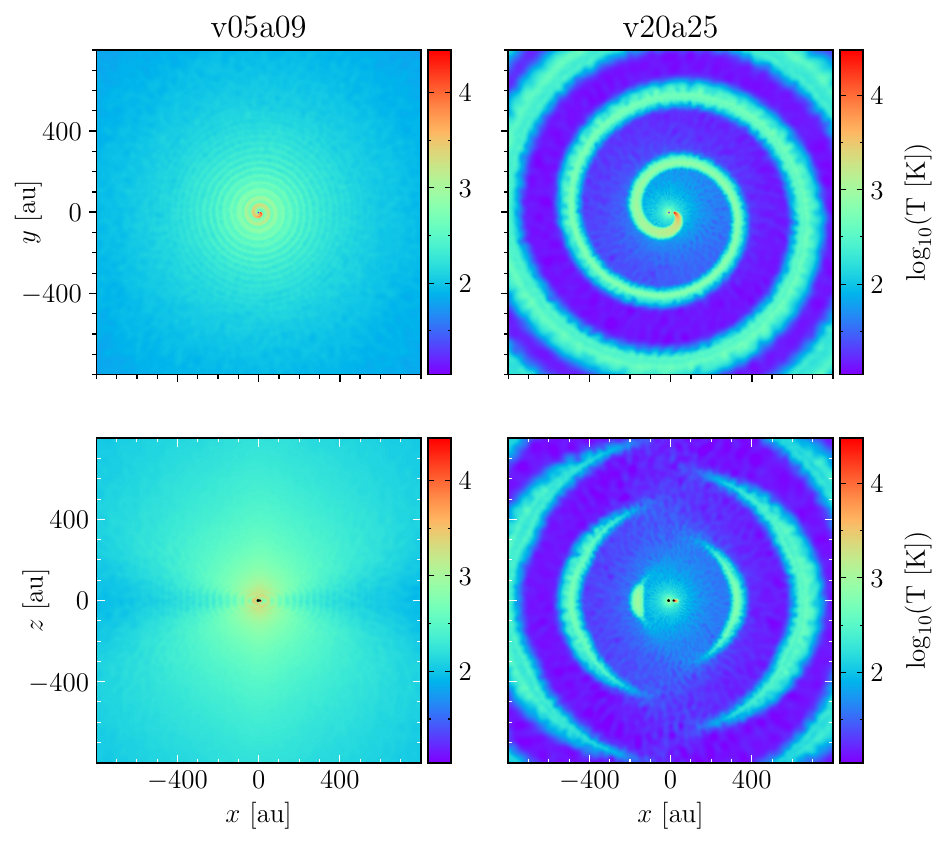}
	\caption{Temperature distributions in slices through the orbital plane (upper row) and the meridional plane (lower row) for the v05a09 (left) and v20a25 (right) models. }
	\label{Fig:tmp}
\end{figure}

For the v20 models viewed edge-on (Fig~\ref{Fig:trans_v20_i90}), we see a distinct behaviour. For the lower transitions ($J = 1{\rightarrow}0$ and $2{\rightarrow}1$), the line shapes are quasi symmetric. However, towards higher transitions, these lines become asymmetric, with a distinct double peak forming for the v20a25 model. These peaks differ in strength, similar to the face-on line profiles of the v05 and v10 models, but at the $J = 3{\rightarrow}2$ transition the red peak dominates, whilst the blue peak is stronger at the $J = 6{\rightarrow}5$ transition. Hence, this feature cannot be explained by the effect described in Sect.~\ref{section:lineshapes}, but rather by the fact that the higher $J$ lines probe different regions for different models. To highlight this effect, we show the temperature distribution in a slice through the orbital and meridional plane in Fig.~\ref{Fig:tmp}, for the v05a09 (left) and v20a25 (right) models, to show two extreme cases. In the v05a09 model, the temperature distribution is quite smooth, with the spiral only barely visible in the innermost region. Higher $J$ lines probe hotter regions, but as the temperature profile is quite uniform, the overall line profile is mainly determined by the global wind morphology. In contrast, the temperature distribution of the v20a25 model is strongly dominated by the spiral structure. This implies that, while the lower $J$ transitions are set by the global morphology, the higher $J$ transitions probe the spiral structure where the temperature is higher. This can produce the strong asymmetries visible in Fig.~\ref{Fig:trans_v20_i90}, as the spiral structure at blue and red shifted velocities differ by half a period. 

In Fig.~\ref{Fig:trans_i0} we show the $J = 1{\rightarrow}0$ and $6{\rightarrow}5$ transitions for the v10 and v20 models, viewed face-on. For the v10 models, the same general results described above apply, as the overall line shape remains largely the same, with some differences in the dip between the two peaks. For the v20 models, the line shapes change significantly, due to the transition from the global wind to the spiral structure. For the $J = 1{\rightarrow}0$ transition, the lines are similar to those in Fig.~\ref{Fig:CO2-1_i=0}, with a more pronounced double peak for the v20a09 model. At the $J = 6{\rightarrow}5$ transition, all lines show two peaks, and most notably, the v20a25 model now has a central bump, not too dissimilar from the edge-on line profiles of the v05 and v10 models. This central bump comes from the fact that the concentric arcs are slightly thicker towards the orbital plane, than towards the polar regions, as visible in the lower right plot of Fig.~\ref{Fig:morphologies_v20}. In the face-on view the spiral structure is symmetric at the blue and red shifted velocities, and as a result, the spectral lines are also quasi symmetric, in contrast with the edge-on view. 

Finally, in Fig.~\ref{Fig:trans_i90} we show the $J = 1{\rightarrow}0$ and $6{\rightarrow}5$ transitions for the v05 and v10 models viewed edge-on. Here, the main features of the line, i.e. the strong central bump, are still present, but the general shape has still changed significantly with $J$. For both v05a09 and v10a09 at the $J = 1{\rightarrow}0$ transition, two small peaks at the outer edges of the line become visible. At the $J = 6{\rightarrow}5$, the central bump has increased in strength relative to the outer wings for all models. The v10a25 model, which does not possess a central bump, shows a very small dip at the centre.
\begin{figure}[t]
	\includegraphics[width = \columnwidth]{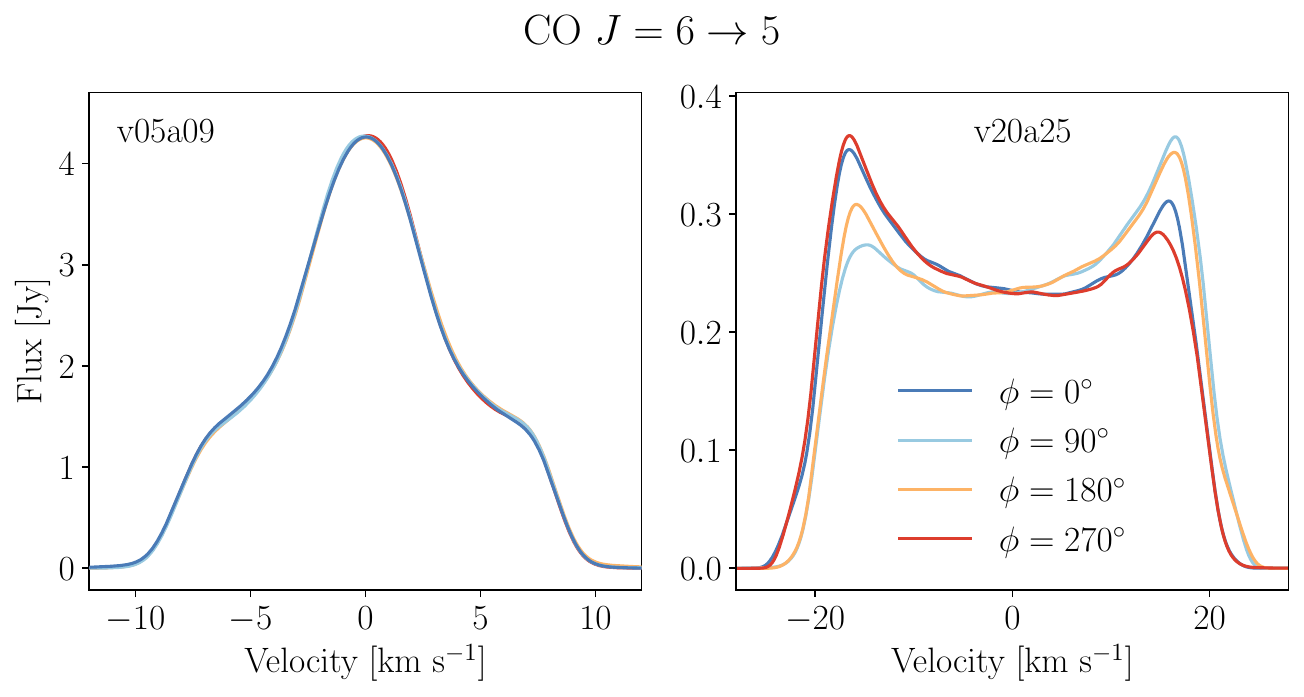}
	\caption{Synthetic CO $J = 6{\rightarrow}5$ spectral lines for the v05a09 (left) and v20a25 (right) models, viewed edge-on and at four different values of the PA ($\phi$). The different colours correspond to the different values of the PA.}
	\label{Fig:PA}
\end{figure}

\begin{figure*}[t]
\centering
	\includegraphics[width =0.965\textwidth]{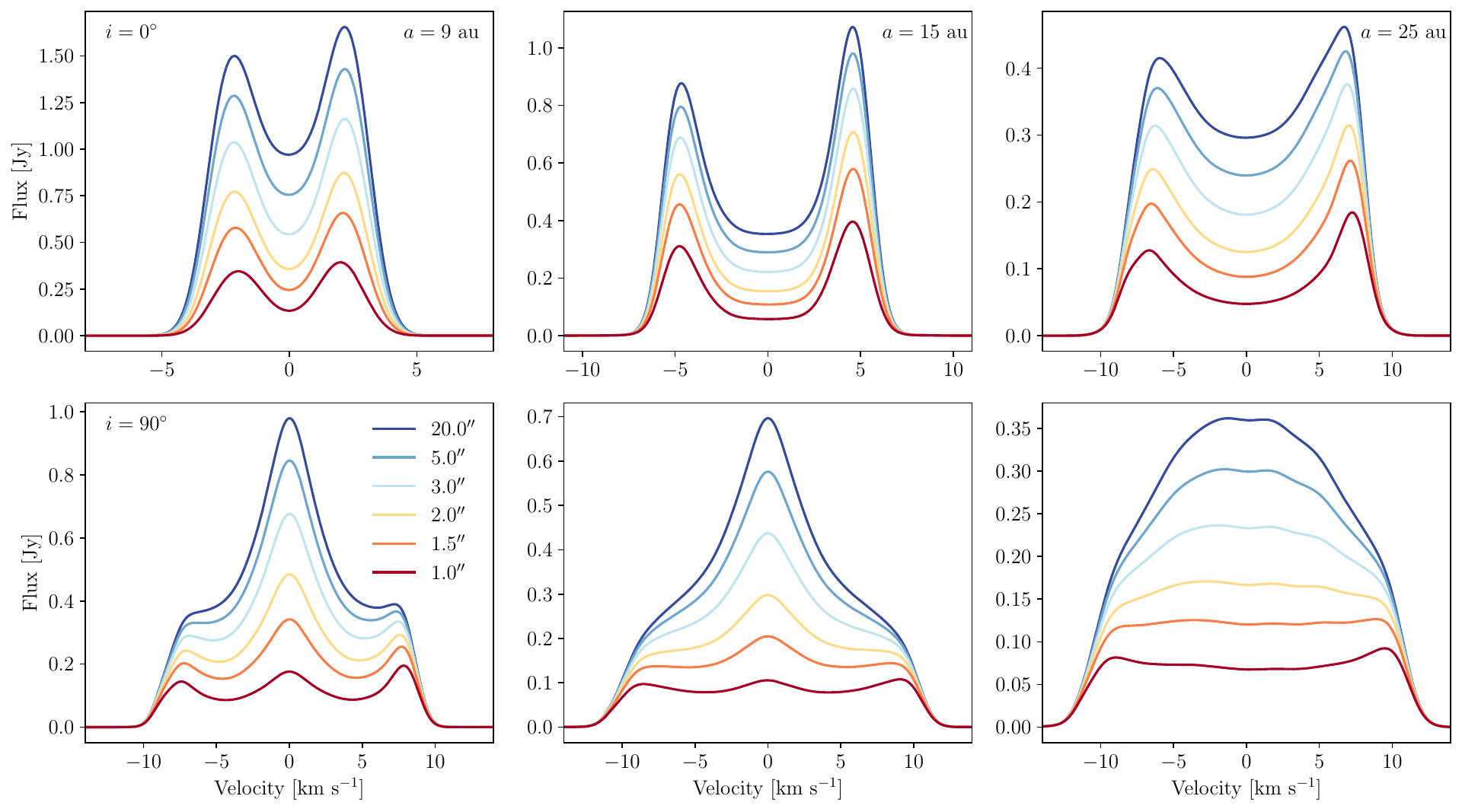}
	\caption{Synthetic CO $J = 2{\rightarrow}1$ spectral lines for the models with initial wind velocities of 10 km s$^{-1}$, where a beam with five different values of $\theta_b$ has been applied. The spectral lines are viewed face-on at an inclination of $0 \degree$ and edge-on at $90 \degree$ in the upper and lower row, respectively. From left to right, the orbital separation is 9, 15, and 25 au.}
	\label{Fig:beam}
\end{figure*}

\subsubsection{Effect of PA}
\label{section:effect of PA}
So far, all the spectral lines have been created at a PA of 0 degrees. Given the results of Sect.~\ref{section:higher co}, we can make a prediction whether the PA will play a role in the line shape as well. As long as the shapes of the line profiles are determined by the global wind, as is the case for the v05 and part of the v10 models, the PA will have a negligible effect. In contrast, as described above the higher $J$ lines for the v20a25 model directly probe the spiral structure. Hence, by changing the PA, the projected velocities of the different parts of the spiral are altered, thereby also changing the spectral line. 

This is illustrated in Fig.~\ref{Fig:PA}, where the $J = 6{\rightarrow}5$ line viewed edge-on is shown, for four different values of the PA, differing by a quarter of a period for the v05a09 (left) and v20a25 (right) models. Changing the PA for the v05a09 has a very small effect on the line profiles, contrary to the v20a25 model where the location and the strength of the stronger peak strongly depend on the PA. This effect is strongest when viewing perfectly edge-on, and decreases in strength towards smaller inclinations, disappearing when viewing face-on. This phase dependency of the spectral line could be a diagnostic for locating the companion. Due to the self-similarity of the models, and low optical depth, spectral lines that are viewed at a PA that differ by $180\degree$ are near perfectly mirrored. This would likely no longer be the case if we introduce eccentricity in our models. 

\subsubsection{Single dish telescopes}
\label{section:beam}
When observing AGB outflows with single dish telescopes, the spectral features can be heavily influenced by the beam profile, depending on whether the source is resolved or not. To see how the spectral lines are influenced by the beam, we apply Gaussian beams with different widths on the CO $J = 2{\rightarrow}1$ line of the three v10 models. Since the model has an outer boundary at 5000 au, and is at a distance of 500 pc, the model spans $20''$. We apply beams, centred on the CoM, with five different beam sizes of $\theta_b \in [20'', 5'', 3'', 2'', 1.5'', 1'']$. These beam sizes are selected purely for visual purposes, and represent different ratios between the beam size and size of the model. This is done for the v10 models, both on the face and edge-on line profiles. 

The resulting line profiles (Fig.~\ref{Fig:beam}) show that when viewing models face-on (top row), the overall line features are preserved, with two distinct peaks present regardless of beam size. This is to be expected given the origin of these peaks. Applying a Gaussian beam generally affects the flux at the central velocities more, as these have a greater extent on the sky. The peaks originate above the CoM (see the top row of Fig.~\ref{Fig:CO2-1_i=0}) at the greatest velocities towards and away from us. Thus, they subtend a smaller area on the sky, which is near the centre of the beam, and  are less affected. Viewing edge-on, the line shapes change quite drastically when varying the beam size. The central peak remains visible for the v10a09 and v10a15 model, although the strength compared to the wings decreases, and at the smallest beam sizes, two small peaks start to form around the central peak for the v10a09 model. This is in line with expectations, as the beam removes more flux from the central velocities. When the beam is very small, this effect can create a double peaked line profile even in spherically symmetric outflows. 

Finally, for the v10a25 model, we observe a similar behaviour, where the two peaks in the face-on view remain very pronounced, even at the smallest considered beam. In the edge-on view, where the line profile starts out as quasi parabolic, the line slowly becomes flat topped, and two peaks start forming at the smallest beam. This is very similar to a spherically symmetric outflow, as only the region in front and behind the star is captured within the beam, and gas moving at larger angles with respect to the line of sight are excluded more.

\section{Discussion and future work}
\label{section:discussion}

\subsection{Implications}
\label{discussion:impl}

Given the strong deviations from flat-topped and parabolic line profiles, direct observations of line profiles as shown in Sect.~\ref{section:results} can be indirect evidence of a hidden binary companion. However, the line shape can be very degenerate with the orbital and wind parameters, as attested by the strong similarities between the v05 models and the v10a09 and v10a15 models. Additionally, all the v20 models show strong similarities for all spectral lines, making it very hard to retrieve the orbital parameters solely using a spectral line. Not only can the line shape change with the orbital parameters, as clearly shown when varying the separation of the v10 models (Fig.~\ref{Fig:CO2-1_i=0}, middle column), but also its strength. This is most visible for the v05 models, whose line shape is similar between the different orbital separations, but the line strength of the $J = 2 {\rightarrow} 1$ transition of the v05a25 model is almost twice as high as those of the v05a09 and v05a15 models (see Fig.~\ref{Fig:CO2-1_i=0}). In addition, the behaviour of the spectral lines when moving to higher $J$ transitions can also differ between models, both in line shape and strength. Hence, observing multiple lines simultaneously can alleviate this degeneracy.

Closely related to this, is the terminal wind velocity one retrieves from the lines, whose value strongly depends on both the orbital parameters, as well as the inclination. For the v05 models viewed face-on, the terminal velocity is approximately the input velocity, though it has slightly decreased for the v05a25 model. When looking edge-on, however, all three considered orbital separations have terminal velocities around 10 km s$^{-1}$. As we do not include radiative acceleration in our models, the acceleration of the material can only be achieved by the interaction with the secondary star. In contrast, all v20 models have terminal velocities close to the input velocity, regardless of inclination and orbital separation, as the interaction with the companion is less strong. For the v10 models, we see a strong dependence on the orbital separation in the face-on view, as the 9, 15, and 25 au models have terminal velocities of approximately 5, 7.5, and 10 km s$^{-1}$, respectively. For the v10a09 and v10a15 models, this is even lower than the initial velocity. In the edge-on view, the difference between the models is smaller, as the v10a15 and v10a25 models have very similar velocities around 12 km s$^{-1}$, and the v10a09 model has 10 km s$^{-1}$. 

\citet{elvire} derived a fitting function (their Eq.~9) for the mass-loss rate, that depends on the integrated line intensity of low-$J$ lines, the terminal wind velocity $v_\infty$, fractional CO abundance, and parameters linked to the spherically symmetric models. Using this expression with typical values relevant for spherically symmetric models, we calculate the mass-loss rate of our models, viewed both face and edge-on, by combining the result of the analytical expression for the $J = 1{\rightarrow}0, 2{\rightarrow}1,3{\rightarrow}2,4{\rightarrow}3,6{\rightarrow}5$, and $7{\rightarrow}6$ transitions. It should be noted that this equation is the result of a minimisation over a grid of spherically symmetric models, where detailed cooling and heating mechanisms have been taken into account. In our hydro models, we have only included H\textsc{i} cooling, which strongly affects the integrated line strengths. Therefore, the exact value retrieved from this formula is not of greatest importance, and we instead wish to see the difference in retrieved mass-loss rate between the models and between inclinations. The input mass-loss rate for all models is $10^{-7}$~M$_\odot$~yr$^{-1}$. For reference, we ran a single star simulation in \textsc{Phantom} using an initial wind velocity of 10 km s$^{-1}$ and a mass-loss rate of $10^{-7}$~M$_\odot$~yr$^{-1}$. Using the formula of \cite{elvire} we obtained a mass-loss rate of  $\sim 5 \times 10^{-8}$~M$_\odot$~yr$^{-1}$, which can be attributed to omitting the cooling and heating mechanisms.  

\begin{table}[ht]
    \caption{\label{tab:massloss} Retrieved mass-loss rates using the analytical expression of \citet{elvire}, viewed face-on and edge-on, in units of $10^{-7}$ M$_\odot$ yr$^{-1}$.}
    \begin{center}
        \begin{tabular}{c c c} 
        \hline
        \hline
        Model name &  $\dot{M}$ [$10^{-7}$ M$_\odot$ yr$^{-1}$]  & $\dot{M}$ [$10^{-7}$ M$_\odot$ yr$^{-1}$]  \\
         & $i = 0\degree$  & $i = 90\degree$ \\
        \hline
         Single & $0.47 \pm 0.10 $ & $0.47 \pm 0.10$ \\ 
         v05a09 & $0.58 \pm 0.08 $ & $1.16 \pm 0.15$ \\ 
         v05a15 & $0.65 \pm 0.10 $ & $1.40 \pm 0.20$ \\
         v05a25 & $0.76 \pm 0.15 $ & $1.60 \pm 0.31$ \\
         v10a09 & $0.61 \pm 0.10 $ & $1.38 \pm 0.22$ \\
         v10a15 & $0.75 \pm 0.10 $ & $1.20 \pm 0.12$ \\
         v10a25 & $0.86 \pm 0.08 $ & $1.12 \pm 0.10$ \\
         v20a09 & $0.85 \pm 0.05 $ & $0.98 \pm 0.07$ \\
         v20a15 & $0.99 \pm 0.07 $ & $1.19 \pm 0.08$ \\
         v20a25 & $1.04 \pm 0.07 $ & $1.14 \pm 0.09$ \\ 
        \hline 
    \end{tabular}
    \end{center}
\end{table}

The derived values for the mass-loss rate of each model, which represent the mean value and standard deviation of all the considered lines, viewed face-on and edge-on are shown in Table.~\ref{tab:massloss}. First of all, the derived values from the binary star simulations deviate from the single star \textsc{Phantom} model, which again can be contributed to a change in temperature profile, in this case caused by heating due to compression by the binary companion. Therefore, the good agreement between input and output mass-loss rates is likely a coincidence. More interestingly, the estimates of the mass-loss rate when viewing face or edge-on can be quite different. For the v05 and v10 models, this difference between the inclinations is typically a factor of two. For the v20 models, the derived mass-loss rates agree within 1$\sigma$ between the face and edge-on views. This difference can be attributed to a change in the estimate for the terminal wind velocity $v_\infty$. When using spherically symmetric models,  $v_\infty$ is strongly connected to the density, and hence to the mass-loss rate. However, when viewing the models at different angles, $v_\infty$ changes as described above, causing a difference in the derived mass-loss rates. The difference between the retrieved mass-loss rates of the single star and binary star simulations shows that even in the case where the line profiles approach a simple flat topped or parabolic shape (i.e. the v10a25 and v20 models), the retrieved mass-loss rate can still differ by at least a factor of two, due to impact the binary companion has on the underlying temperature profile. 

\subsection{Observability of line shapes}
\label{section:noise}
\begin{figure*}[ht]
\centering
	\includegraphics[width =0.965\textwidth]{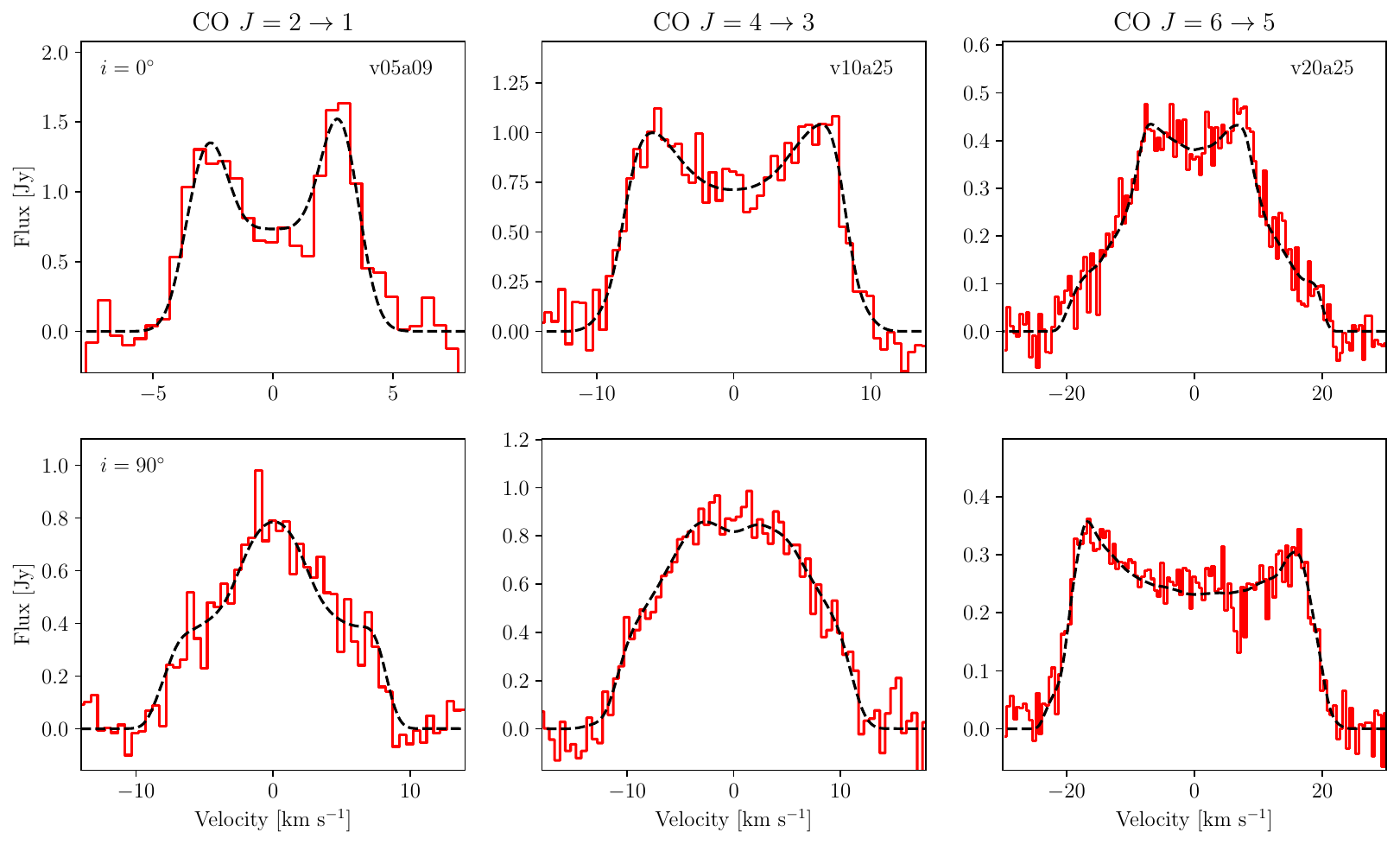}
	\caption{Synthetic CO spectral lines for three selected models. From left to right these are v05a09, v10a25, and v20a25 with the $J = 2{\rightarrow}1, 4{\rightarrow}3, 6{\rightarrow}5$ transitions, respectively. The spectral lines are created viewing face-on (upper row) and edge-on (lower row). The black dashed lines show the original spectral line, and the red line has the addition of white noise assuming a S/N of 10, and an increased velocity bin size to 500 m s$^{-1}$.}
    \label{fig:noise}
\end{figure*}

When observing CO spectral lines, one also has to worry about noise present on the observations. Additionally, we have worked at a spectral resolution of 0.1 km s$^{-1}$ so far; however, this can be significantly larger for observations, although single dish telescopes can reach high spectral resolutions. One can then wonder which of the features described above remain observable. A detailed description of these effects is out of the scope of this paper, and instead we increase the velocity bin size to 0.5 km s$^{-1}$, and add white noise on top with a $1\sigma$ uncertainty of 10\% of the peak flux, corresponding to a signal-to-noise (S/N) of 10. This is shown in Fig.~\ref{fig:noise}, where we show the CO $J = 2{\rightarrow}1, 4{\rightarrow}3$ and $6{\rightarrow}5$ lines for the v05a09, v10a25 and v20a25 models, respectively, both viewed face-on and edge-on. The underlying real spectral line is always shown as a dashed black line. 

First, looking at the v05a09 model, generally the two distinct features of the spectral lines remain distinguishable. Especially the double peaks in the face-on view remain well visible. The central bump feature in the edge-on view is also visible, although the black dotted line helps significantly in identifying it, and it can be easily misinterpreted as a triangular line profile, as the transition from the outer region of the spectral line, to the central bump is easily hidden by the noise. The double peaks of the v10a25 model also remain visible, though they have become less pronounced, and in some case could disappear. The small features on the quasi parabolic profile in the edge on view have been entirely encapsulated within the noise. Similarly, all the small features that are visible on the lower $J$ lines of the v20 models are also likely to be hidden within the noise. At the highest transitions, some features become visible for the v20a25 model. In the face-on view, one can still distinguish the central bump. The two peaks present on this central bump are still present, although without the knowledge of the underlying spectral line, this goes unnoticed. The double peaks, whose strength depends on the PA, remain visible to a certain extent, though it is unlikely that any changes in the shape due to the PA will remain significant for our models. Fig.~\ref{fig:noise_20} shows the same spectral lines, but now assuming a S/N of 20. Immediately, all the characteristic features become more pronounced. Most noteworthy is that for this S/N, the asymmetric peak for the $J = 6{\rightarrow}5$ transition of the v20a25 model, likely will remain observable.

On top of the noise, we need to consider the beam profile as done in Sect.~\ref{section:beam}. In the case where the emitting region is larger than the beam, the change in line shape combined with the noise will make it very difficult to retrieve the true underlying line shape, and hence infer any information about the orbital parameters. This is especially true for any double peaked patterns, as these can originate from a spherical outflow if the outflow is resolved, and thus can be easily miscategorised. This degeneracy could be lifted by including additional observables (e.g. channel maps), but this is out of the scope of the current study, and barely available for a large sample of stars.

\subsection{Comparison with earlier studies}
Given the difference betweens the methods used in similar studies, it is difficult to directly compare the results. Still, we can see how our overarching results fit in. Although all our models possess a clear spiral structure, our results deviate strongly from the analytical spiral models presented in \citet{Ward2}. In this paper, the authors distinguish between narrow and shell spirals (i.e. small and large opening angles), which are embedded in a spherical outflow. Therefore, any flattening effects, as present in the v05 and v10 models, are omitted. Furthermore, they impose a temperature profile on the spiral, and thus neglect the heating by compression caused by the companion. For the narrow spiral models, they obtain spectral lines with a central bump feature, and double peaked line profiles, similar to the spectral lines presented here, although both features occur at different inclinations compared to ours, with the bump feature occurring in the face-on view, and the double peaks in the edge-on view. The origin of these line shapes is very distinct, easily explaining this discrepancy. For the shell spiral, the deviation of their spectral lines from a simpler flat topped or parabolic line profile is less pronounced but still present for some of the models. Our v20 models are representative of the shell spirals, and the lines show a similar deviation from the spherically symmetric case. 

\citet{Kim_2019} study four HD simulations which cover a distinct part of the parameter space compared to our models. The models have an orbital separation of 68 au, two values for the eccentricity are considered (0 and 0.8), and the gravitational influence of the companion for two of their models is turned off to isolate the effect of the motion of the AGB star around the CoM. Additionally, the spectral lines are created by calculating the column density within velocity bins, thereby omitting any optical depth or temperature dependence effect. In general, the authors find double peaked profiles when viewing face-on, not too dissimilar from the low-$J$ lines of the v20 models. These are, however, also strongly present when viewing edge-on. We do recover double peaks in the edge-on view for the v20 models when looking at higher $J$ lines (Fig.~\ref{Fig:trans_v20_i90}), but the lines of \citet{Kim_2019} are representative of low-$J$ optically thin lines given method they use, and so do not correspond to any of our results. A more in depth comparison will not be instructive, given the difference in method and orbital parameters. 

\subsection{Comparison with observations}

Although a one to one comparison between observations and our simulations is not within the scope of this paper, we can still compare the global trends observed. First, the strength of the lines is comparable with observations, leading to a general agreement of the mass-loss rates in Sect.~\ref{discussion:impl}. A direct comparison can only be made with oxygen rich targets whose mass-loss rates are close to our input mass-loss rate of $10^{-7}$ M$_\odot$ yr$^{-1}$. To this end, we can compare to W~Hya, T~Mic, Y~Scl, V1943~Sgr and BK~Vir from \citet{2020A&A...640A.133R}, who have mass-loss rates ranging between $8\times10^ {-8}$ and $1.5\times10^ {-7}$ M$_\odot$ yr$^{-1}$, with terminal velocities around 5 km s$^{-1}$. After adjusting the observations for the difference in their distance compared to our models (our models are at 500 pc), we find peak flux values for the $J = 2 {\rightarrow} 1$ and $3{\rightarrow}2$ transitions in the range of 0.5 to 2 Jy, and 1 to 3.5 Jy, respectively. Given the terminal velocity of the observations, we can compare them to the peak fluxes of the v05 models, which fall in the same range as the observations (see Figs.~\ref{Fig:trans_i0} and \ref{Fig:inclinations_v05}). The observations in \citet{2020A&A...640A.133R} were made using ALMA, and so there will be resolved out flux, although this should remain minimal in the used configuration.  

Moving to the line shapes, most striking is the appearance of the central bump feature, which is observed in targets such as RS Cnc, X Her, AFGL 292, V370 And, and EP Aqr \citep{success, 2019A&A...629A..94D}. We  note, however, that EP Aqr has been identified as a face-on narrow spiral in \citet{Epaqr, 2020A&A...642A..93H}. This is in contrast with our results, where the central bump feature is observed when viewing edge-on. The authors derive a separation of $\sim 65$ au, and a companion mass of 0.1 M$_\odot$, which is well outside the parameter space covered by us. Additionally, these types of spectral lines have been modelled in \citet{2019A&A...629A..94D} using an oblate spheroid attached to a biconical structure, viewed nearly face-on. Our models where global flattening occurs do resemble oblate spheroids, but we recover the central feature in the edge-on view. This as a result of the impact of the binary companion on the velocity field, as more material has built up in higher density regions near the central velocities.

A second similarity are the double peaks formed when observing some models face-on. Care has to be taken, since as discussed in Sect.~\ref{section:beam}, double peaks can also form as a result of the beam profile. In \citet{2019A&A...629A..94D}, such double peaked line profiles were again modelled using an oblate spheroid in \citet{2019A&A...629A..94D}, now viewing edge-on. This is again in contrast with our models, as we recover peaks when viewing at a small inclination, as the origin of the peaks is the additional region towards the $z$-axis (see the arc-like structures in Fig.~\ref{Fig:velocities_v05}). Other targets show also very pronounced double peaks, although in some cases these are known to originate from detached shells, for example R Scl \citep{R_scl1, R_scl2}. We cannot yet produce such shells in our simulations, as these are believed to be caused by a brief increase in the mass-loss rate during a thermal pulse.

Besides these stronger features being observed in some systems, some additional subtle results are also retrieved. The emergence of a double-peaked pattern at higher $J$ transitions is observed in V Aql, V CrB and W And \citep[see Fig.~A.1 and Fig.~A.2 in][]{success}. These line profiles also become more asymmetric towards the higher transitions, similar to our results of Sect.~\ref{section:higher co}. A single peak to the side of the central velocity, similar to the $J = 6{\rightarrow}5$ transitions of the v05a25 model as seen in Fig.~\ref{Fig:trans_v5_i0}, can be seen in R Cas \citep[see Fig.~2][]{elvire}. However, to be similar in nature, we expect the lines to be reasonably optically thick. Smaller scale deviations from flat topped or parabolic lines, similar to the v20 models (see Fig.~\ref{Fig:CO2-1_i=0}) are also often observed, for example in SW Vir, RX Boo, R And, and RT Vir \citep{success}.  

\subsection{Limitations and future work}

It is important to keep in mind the assumptions made in both the \textsc{Phantom} models and the post-processing with \textsc{Magritte}, and the impact these have on our results. As noted in Sect.~\ref{section:method}, we neglect the radiative acceleration in our models, and instead opt for a free wind approach. This implies that the velocity profile in our simulations is also unrealistic. As shown in \cite{SM1}, the wind velocity at the location of the companion strongly impacts the complexity of the resulting morphology, and so obtaining a realistic velocity profile is crucial. In order to further refine the wind launching mechanism, we need to consider detailed dust nucleation \citep[as done in][]{siess} and radiation pressure \citep[as done in][]{esseldeurs}, though this still needs to be combined with pulsations to obtain physically correct density structures in the inner region, which is also tied to the dust formation. Therefore, we aim to implement a more accurate radiative pressure formalism, and pulsations into \textsc{Phantom}, similar to \cite{aydi_mohamed}. 

Currently, all our models show very pronounced spiral like structures. Although some observations indicate a well-defined spiral structure \citep[see e.g. AFGL 3068 in][]{2006A&A...452..257M, 2006IAUS..234..469M}, the majority reveal more chaotic patterns can exist.  Including eccentricity is one pathway to introduce more asymmetry \citep{JM2}. However, detailed 3D RHD simulations of AGB stars, where pulsations and convection emerge self-consistently \citep[see e.g.][]{Freytag_2023, self_excited} show that mass loss can be non-spherically symmetric due to large convection cells and non-radial pulsations. Implementing anisotropies in the particle ejection could have an appreciable effect on the complexity of the overall morphology. 

Additionally, we do not yet include detailed cooling mechanisms \citep[see e.g. the mechanisms described in][]{2006}. The effect on the line shapes are expected to be minor, although the line strengths can deviate very strongly with changing temperature, and greatly impact any mass-loss rate retrieval, as discussed above. Hence, we aim to incorporate more cooling mechanisms in our simulations (Dionese et al. In prep). To obtain the most accurate cooling rates, we need detailed chemistry. The first step has been achieved in \citet{Maes_2024}, who created \textsc{MACE}, a chemistry emulator, which will be coupled to \textsc{Phantom} in the future. 

Neglecting dust has a second effect on our results, as it also acts as an additional source term in the radiative transfer. Most importantly, dust can affect the level populations by exciting CO to its first vibrational transition, leading to IR pumping \citep{Lamers_Cassinelli_1999}. By including the first vibrational level of CO, the amount of transitions that need to be considered increases by a factor of four, leading to a significant increase in computational cost of the radiative transfer. Furthermore, as shown in \citet{Matsumoto_2023} dust can also attenuate the CO emission, though they find this plays a larger role for higher transitions. Taking into account dust in the radiative transfer is again not expected to significantly alter the line shapes, but mainly impact the line strengths.

It is important to stress that a very small part of the possible parameter space has been covered, and that a great number of relevant parameters are taken constant over the grid, most importantly the mass-loss rate, which we justified in Sect.~\ref{section:gridsetup}. Increasing the mass-loss rate will greatly alter the photodissociation radius and line strengths, but given the lack of additional cooling and a detailed description of radiation pressure in the current set-up, the impact on the underlying morphology is expected to remain small. Besides that, we only considered oxygen-rich AGB stars by imposing the corresponding fractional abundance of CO. Increasing this value will strongly alter the photodissociation radius, optical depth for the radiative transfer, and line strengths.

This study focusses on the spectral lines, which is the most often observed feature from AGB outflows. However, a great deal of information can easily be concealed, either in the intrinsic shape of the line or by the noise and beam profile. With the emergence of high-resolution interferometric observations, we can also study different observables, such as channel maps and position-velocity diagrams. This has been studied in \citet{Ward2, Ward1}, but again with the more simplistic analytical models.

\section{Summary and conclusions}
\label{section:conclusion}

Using the SPH code \textsc{Phantom}, we created a grid of nine 3D models of AGB stars with a stellar companion, with different orbital separations and initial wind outflow velocities. The simulations shows distinct AGB outflows exhibiting deviations from spherical symmetry, both in the inner region, where a spiral structure arises in the orbital plane, and in the global morphology where flattening can occur. We implemented a novel method to take into account CO photodissociation in complex 3D morphologies that utilises a ray tracing scheme combined with nearest ray interpolation. 

Utilising the 3D NLTE line radiative transfer code \textsc{Magritte}, we computed the level populations of CO considering only rotational transitions. We created synthetic spectral lines of the six lowest transitions of CO for all our models viewed at different inclinations and position angles. A variety of line shapes can emerge, depending on the different input parameters. These includes lines that deviate strongly from the single star scenario, with two strong peaks near the terminal wind velocity when viewed face-on, or pronounced bumps near the central velocity when viewed edge-on. In contrast, quasi parabolic or flat-topped profiles can emerge, where the impact of a companion goes unnoticed. The line strengths, and the evolution of the line when looking at different transitions, or at different values for the PA, can provide useful information on the underlying morphology. 

We mimicked the effect of observing the models with single-dish telescopes in two ways, by applying a beam profile, and by adding noise to the spectral lines. This showed that many of the features of the spectral lines that are characteristic of a binary companion can be easily hidden by these instrumental effects, with the possible danger of misinterpreting the line profiles as originating from spherically symmetric outflows. This shows the difficulty of obtaining information about the system when only considering the spectral lines. Even so, the companion can have a significant influence on the temperature profile, and hence on the level populations of the model, thereby impacting the retrieved mass-loss rate.

Given the sensitivity of the spectral lines on the stellar and orbital input parameters, the lack of detailed cooling and heating mechanisms, and insufficient wind driving physics, a direct comparison of our synthetic spectral lines to observations is not yet possible. We aim to improve upon these shortcomings in the future and to move towards this goal.

\begin{acknowledgements}
OV acknowledges funding from the Research Foundation - Flanders (FWO), grant  1173025N. ME acknowledges funding from the FWO research grants G099720N and G0B3823N. TC acknowledges funding from the Research Foundation - Flanders (FWO), grant 1166724N. LS is a senior researcher for the FNRS. KM acknowledges funding from the Research Foundation - Flanders (FWO), grant 1169822N. FDC is a Post-doctoral Research Fellow of the Research Foundation - Flanders (FWO), grant 1253223N. TD is supported in part by the Australian Research Council through a Discovery Early Career Researcher Award (DE230100183). CL acknowledges support from the KU Leuven C1 excellence grant BRAVE C16/23/009. LD acknowledges support from the KU Leuven C1 excellence grant BRAVE C16/23/009, KU Leuven Methusalem grant SOUL METH/24/012, and the FWO research grants G099720N and G0B3823N.  
\end{acknowledgements}
\bibliographystyle{aa} 
\bibliography{bibl.bib} 
\appendix

\section{Comparison with \textsc{SKIRT}}
\label{appendix D}

To further validate and check the accuracy of our results, we compare them by generating spectral lines with \textsc{SKIRT}\footnote{\url{https://skirt.ugent.be/root/\_home.html}} \citep{skirt}, an open-source 3D Monte Carlo radiative transfer code. Although \textsc{SKIRT} is mainly used for dust radiative transfer, it allows for the detailed NLTE calculations necessary for line radiative transfer \citep{baes, Matsumoto_2023}.

\begin{figure}[htbp]
	\includegraphics[width = \columnwidth]{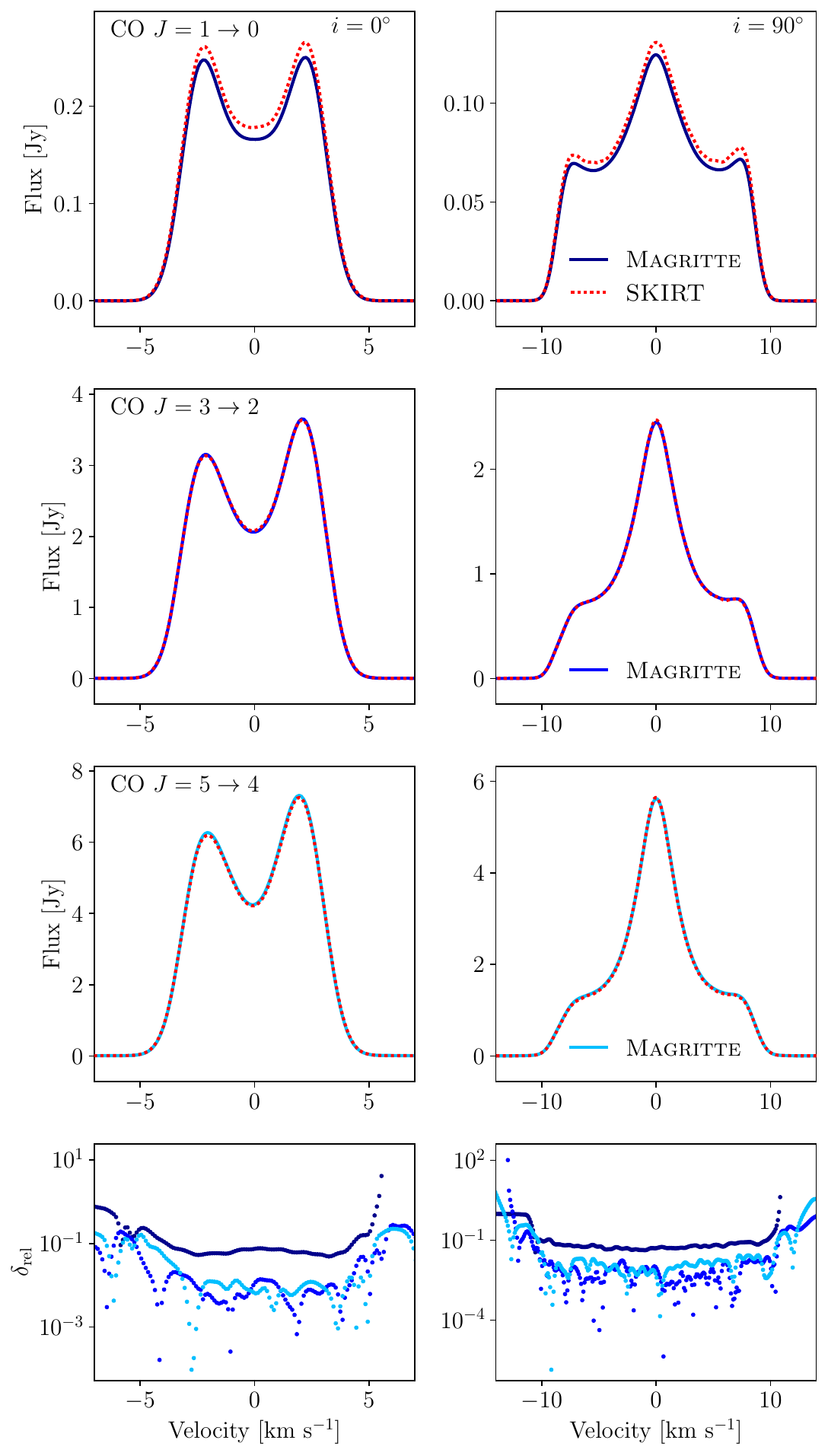}
	\caption{Comparison of the CO $J = 1{\rightarrow}0, 3{\rightarrow}2$, and $5{\rightarrow}4$ lines of the v10a09 model, between \textsc{Magritte} (different shades of blue) and \textsc{SKIRT} (red) for the face-on (left) and edge-on (right) views. The bottom plot shows the relative difference between the spectral lines.}
	\label{Fig:SKIRT}
\end{figure}

We carry out the comparison using the v10a09 model. The results are presented in Fig.~\ref{Fig:SKIRT}, where we show both face-on and edge-on line profiles of the CO $J = 1{\rightarrow}0, 3{\rightarrow}2$ and $5{\rightarrow}4$ lines using \textsc{Magritte} (different shades of blue for different transitions) and \textsc{SKIRT} (red). In the bottom plot we show the relative difference between the line profiles for the different transitions, calculated with Eq.~\ref{eq:reldiff}, where the spectral line from \textsc{Magritte} is taken to be exact. We see that for both the $J = 3{\rightarrow}2$ and $5{\rightarrow}4$ lines, we find nearly perfect agreement, as the relative difference is around one percent for the entire spectral line. Near the wings the difference increases due to the lower values for the flux in that region.

For the $J = 1 {\rightarrow} 0$ line, \textsc{SKIRT} has a systematically higher flux than \textsc{Magritte} for both the face and edge-on views, where the line profiles differ by $\sim 5 - 10\%$, which is still comparable with the expected uncertainty on the lines from observations. Again, the relative error increases towards the wings, due to the lower flux. The line shape still remains the same between the two codes even for the $J = 1{\rightarrow}0$ line. It should be noted that \textsc{SKIRT} was unable to converge completely, which is believed to be linked to this discrepancy. A difference is also present at the $J = 2{\rightarrow}1$ line, but to a lesser extent. However, the agreement between the two codes is very good in general, especially towards higher transitions. 

\clearpage

\section{CO photodissociation}
\label{appendix A}
In this section we explain the implementation of CO photodissociation, which is inspired by \citet{Groenewegen_2017}, and comment on the accuracy of the interpolation scheme. 
\subsection{Implementation}
Our implementation works as follows:
\begin{itemize}    
    \item Starting from the surface of the star, we trace a pre-determined amount of rays outwards, until the edge of the simulation is reached. These rays are chosen such that they are uniformly distributed in 3D. This is accomplished using the HEALPix\footnote{ \url{https://healpix.sourceforge.io/}} package \citep{healpix}, which discretises a sphere into pixels of equal area. The unit vectors, or directions of the rays, are such that they point towards the centre of these pixels. We use HEALPix order of 4, corresponding to 3072 rays.
    \item Along each ray, at all points along that ray \citep[which are determined in a similar way as the ray tracing scheme of][]{Magritte1}, we calculate the H$_2$ and CO column densities by tracing rays in all directions, to the edge of the simulation. These rays are again distributed using HEALPix. In this case, we use HEALPix order 1, corresponding to 48 rays for each point.
    \item Using the calculated column densities, we derive the photodissociation rate using Eqs.~\ref{eq:Ir} and \ref{eq:kr}. When the value of $I(r)$ is known at all the points along a ray, we compute $x(r)$ along that ray using Eq.~\ref{eq:sol}. 
    \item Once this is known for all the initially traced rays, we can interpolate between these to obtain the value of $x_i$ for any other point $i$ of the simulation. This is done by finding the eight closest rays to the point $i$, and interpolating the value of $x$ on each neighbouring ray at the distance $r_i$. The final value of $x_i$ at position $r_i$ is then calculated by using 
    \begin{equation}
        x_i = \frac{1}{\sum_{j=1}^8 d_j^{-2}} \ \sum_{j=1}^8\frac{x_j}{d_j^2},
    \end{equation}
    where $x_j$ is the interpolated value along each ray $j$, and $d_j$ represents the perpendicular distance between the point $i$ and the ray $j$, i.e. closer rays have a larger weight.
    \item We repeat this process, until the relative difference between subsequent iterations becomes less than 0.1 percent, which is typically achieved in five iterations.
\end{itemize}
\subsection{Validation of interpolation}
\begin{figure}[ht]
	\includegraphics[width = \columnwidth]{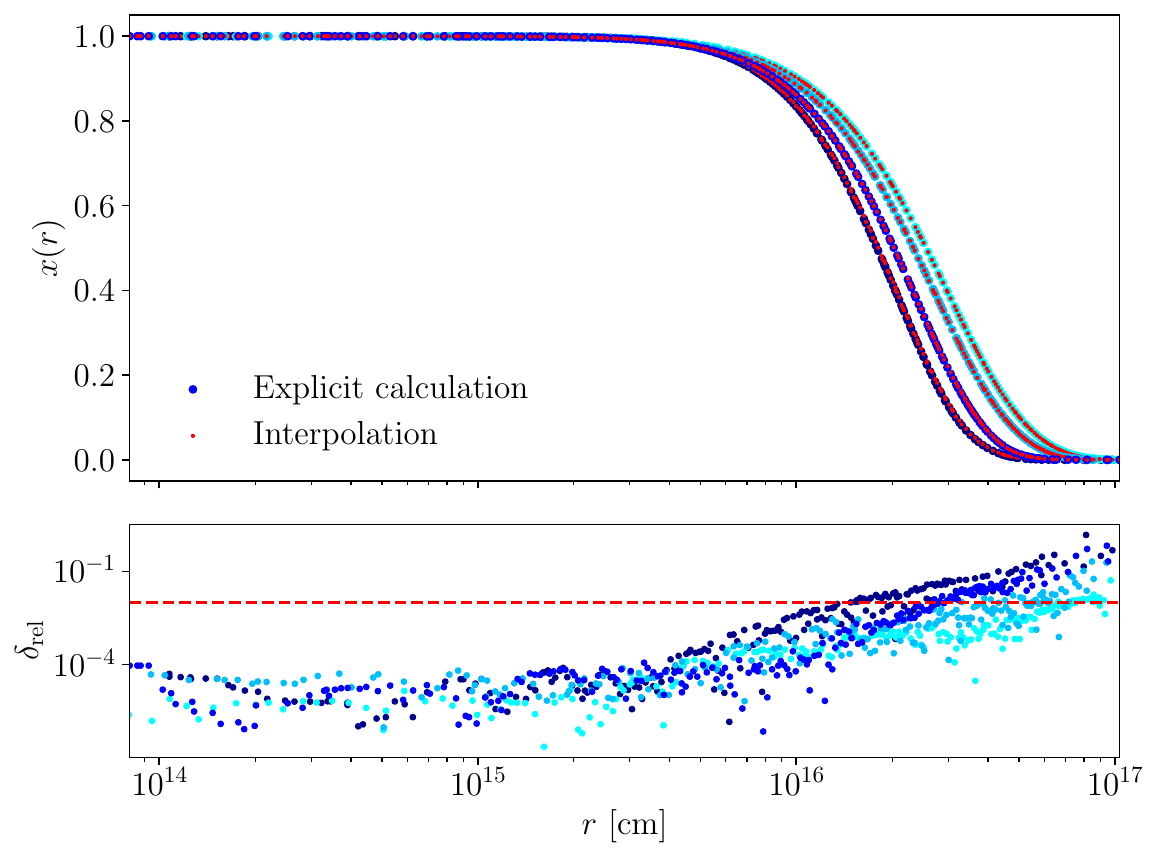}
	\caption{Top: $x(r)$ for rays in four different directions in the v10a15 model. The blue dots show the explicit calculation of $x(r)$, and red the interpolation between the eight nearest rays. Bottom: Relative difference between the explicitly calculated and interpolated values. The red dashed line corresponds to a relative difference of 1\%.}
	\label{Fig:raytracer}
\end{figure}

In order to justify the interpolation approach, we trace rays in four random directions in the simulation, calculate $x(r)$ explicitly using Eqs.~\ref{eq:sol}-\ref{eq:kr}, and also compute $x(r)$ using the interpolation scheme outlined above. This test has been carried out for the v10a15 model, and the result is shown in Fig.~\ref{Fig:raytracer}, where the top plot shows $x(r)$ for the four rays considered, with the blue dots corresponding to the explicit calculation, with different shades showing different directions, and the red dots shows the values of the interpolation. The bottom plot shows the relative difference between the explicit and interpolated values at each point, calculated as
\begin{equation}
\label{eq:reldiff}
    \delta_\mathrm{rel} = \frac{| x_\mathrm{ex}(r) - x_\mathrm{in}(r) | }{x_\mathrm{ex}(r)},
\end{equation}
with $ x_\mathrm{ex}(r)$ the explicitly calculated values and $ x_\mathrm{in}(r)$ the interpolated values. We see that in general, the red dots overlap with the blue dots, showing the interpolation is accurate. The bottom plot shows that the relative error remains under 1 percent for the majority of the points. Only towards the outer regions does the relative difference start increasing, which is due to the decrease of $x(r)$ itself. This region is also less important as it does not contribute to the line profiles, since $x(r)$ is close to zero.

Increasing the number of rays that are traced, either in the amount of rays that can be used for interpolation, or in the number of rays used to calculate the column densities,  improves these results marginally, but at an additional computational cost. We did not find sufficient improvement in the accuracy to justify this increase in time.

\section{Morphologies and velocities}
\label{Appendix B}

This section contains additional slices through the orbital and meridional plane for the v05 and v20 models, for both the density and velocity structures, analogous to Figs.~\ref{Fig:morphologies_v10} and \ref{Fig:velocities_v10} for the v10 models.

\begin{figure*}[t]
	\includegraphics[width = 0.93\textwidth]{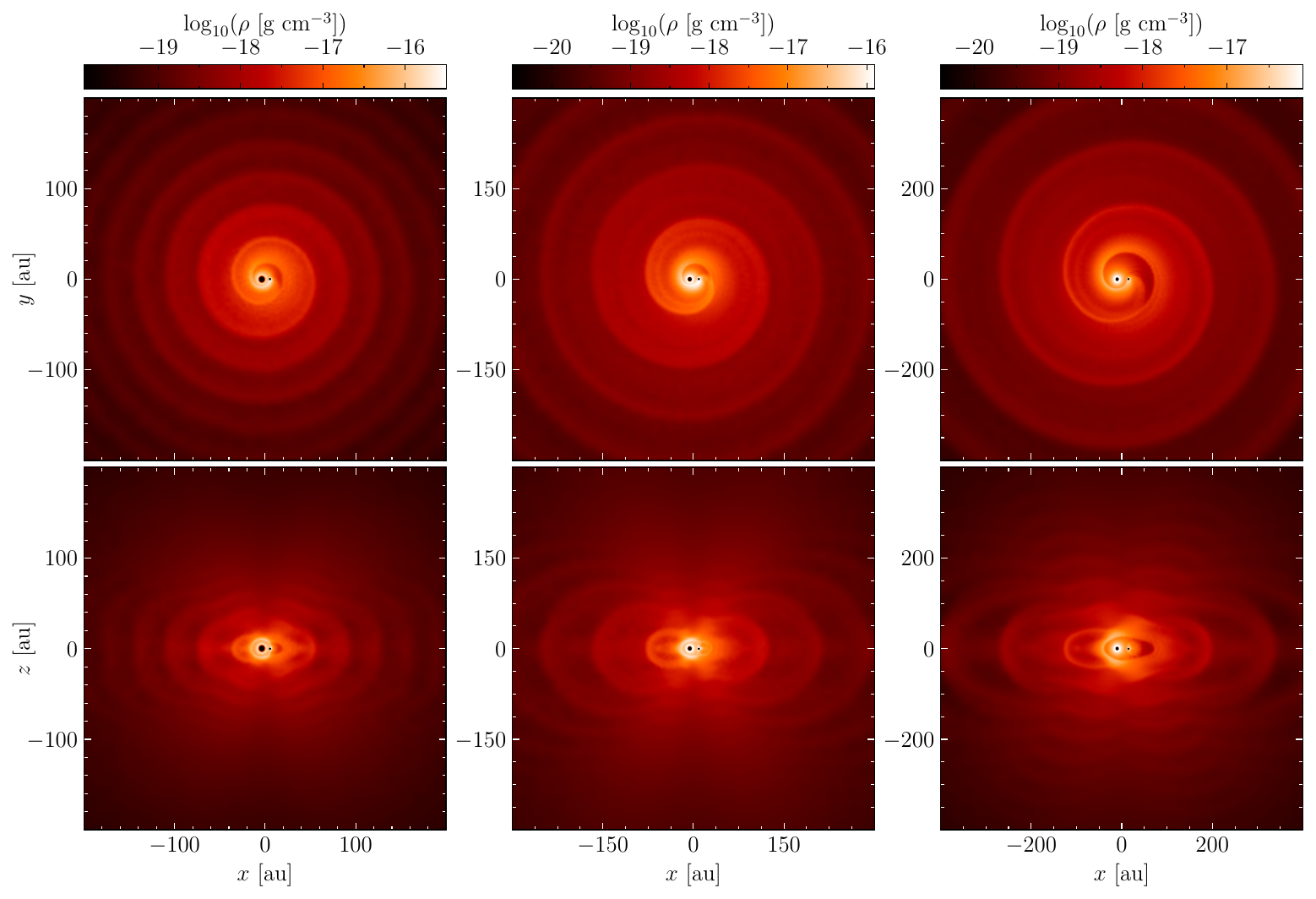}
    \caption{Same as Fig.~\ref{Fig:morphologies_v10}, but for the models with initial wind velocities of 5 km s$^{-1}$.}
    \vspace{-5pt}
	\label{Fig:morphologies_v05}
\end{figure*}

\begin{figure*}[ht]
	\includegraphics[width = 0.93\textwidth]{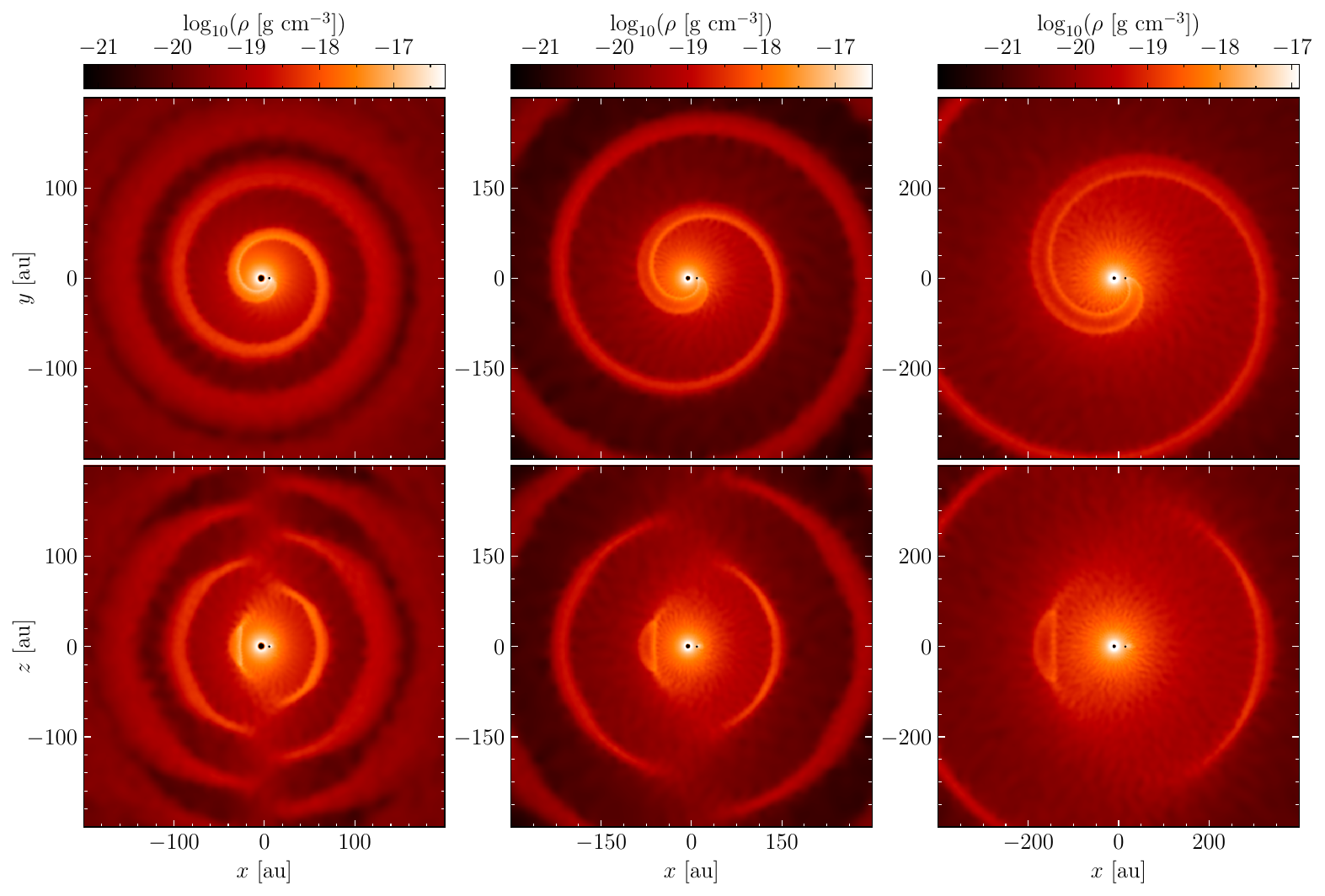}
    \caption{Same as Fig.~\ref{Fig:morphologies_v10}, but for the models with initial wind velocities of 20 km s$^{-1}$.}
	\label{Fig:morphologies_v20}
\end{figure*}

\begin{figure*}[t]
	\includegraphics[width =0.965\textwidth]{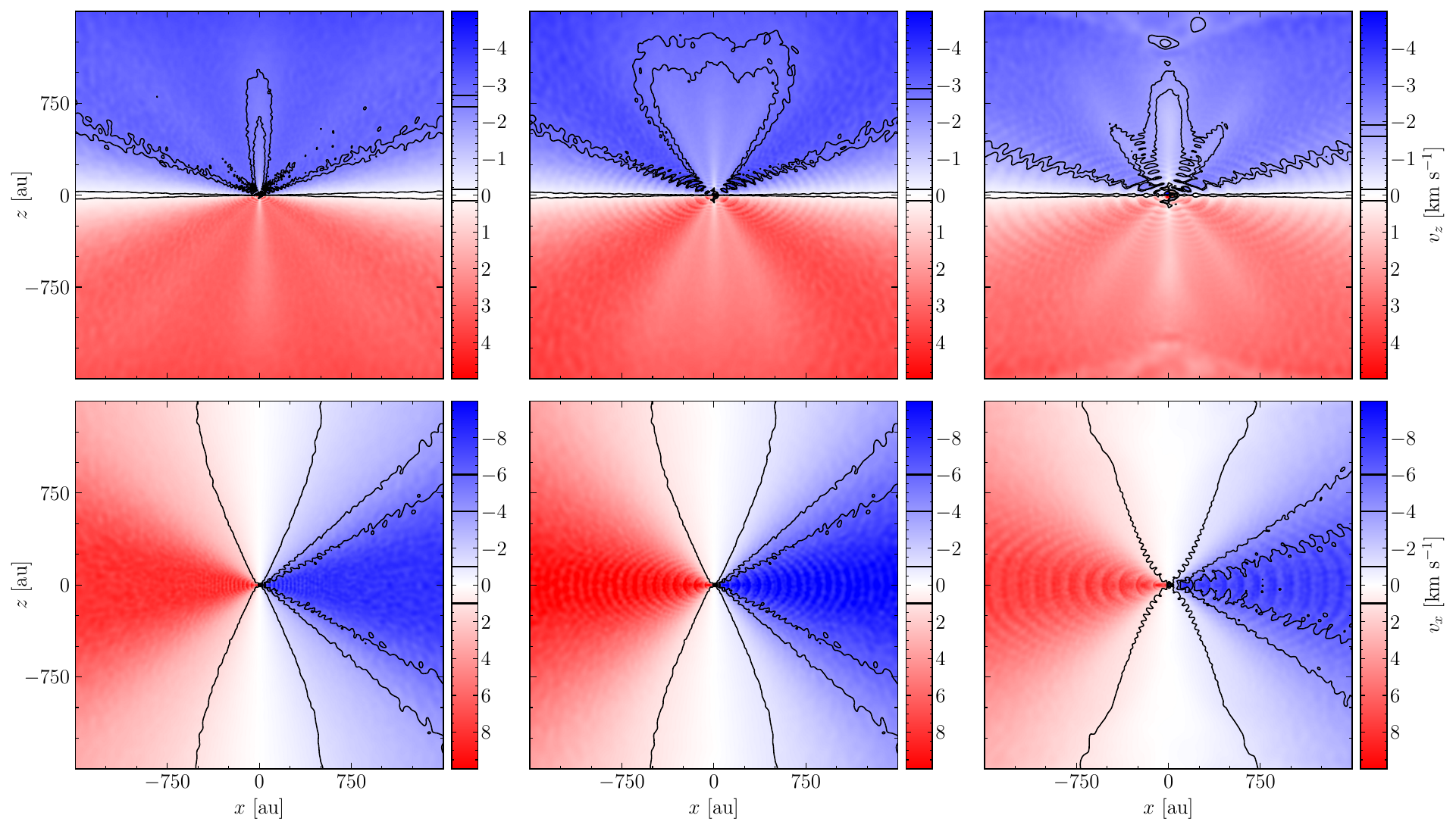}
    \caption{Same as Fig.~\ref{Fig:velocities_v10}, but for the models with initial wind velocities of 5 km s$^{-1}$.}
	\label{Fig:velocities_v05}
    \vspace{-5pt}
\end{figure*}

\begin{figure*}[ht]
	\includegraphics[width =0.965\textwidth]{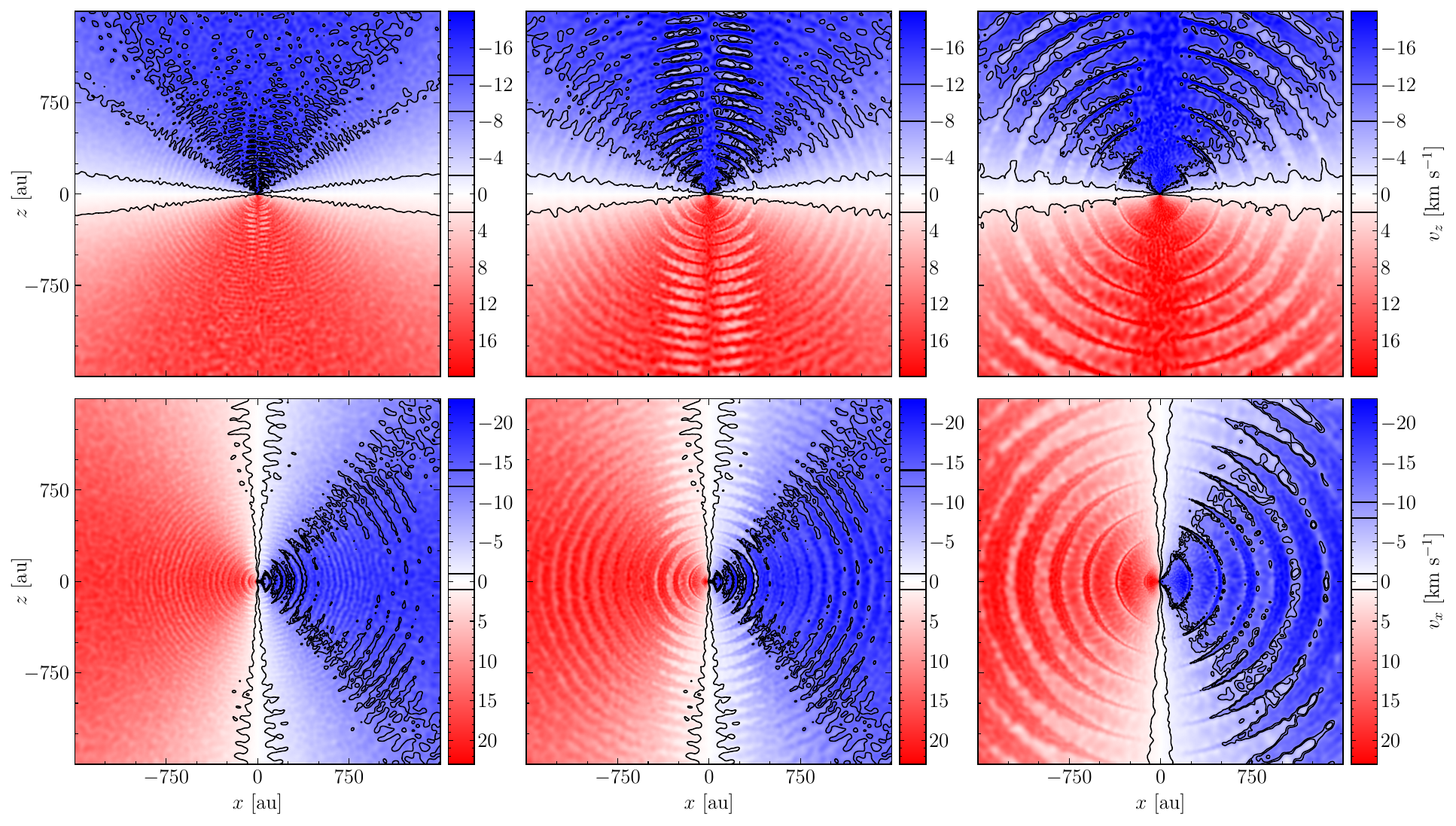}
    \caption{Same as Fig.~\ref{Fig:velocities_v10}, but for the models with initial wind velocities of 20 km s$^{-1}$.}
	\label{Fig:velocities_v20}
\end{figure*}
\clearpage

\clearpage
\section{Additional line profiles}
\label{appendix C}
\subsection{Intermediate inclinations}
This section contains additional plots, showing the synthetic CO $J = 2{\rightarrow}1$ line under intermediate inclinations for the v05 and v20 models, analogous to Fig.~\ref{Fig:inclinations_v10} for the v10 models.

\begin{figure}[ht]
    \begin{minipage}[ht]{\columnwidth}
        \centering
        \includegraphics[width=\columnwidth]{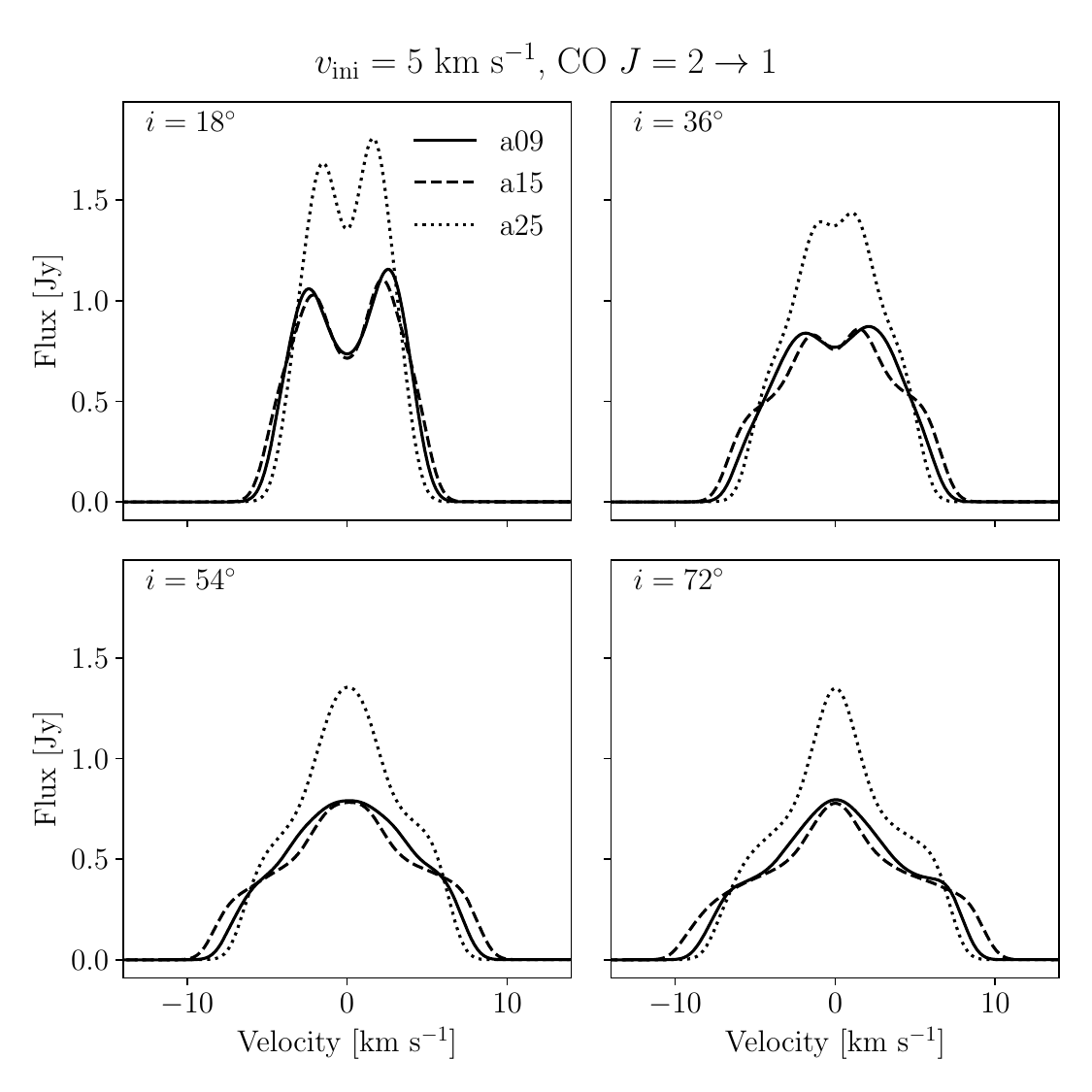}
        \caption{Same as Fig.~\ref{Fig:inclinations_v10}, but for the models with initial wind velocities of 5 km s$^{-1}$. }
        \label{Fig:inclinations_v05}
    \end{minipage}
    \vspace{5mm}
    
    \begin{minipage}[ht]{\columnwidth}
        \centering
        \includegraphics[width=\columnwidth]{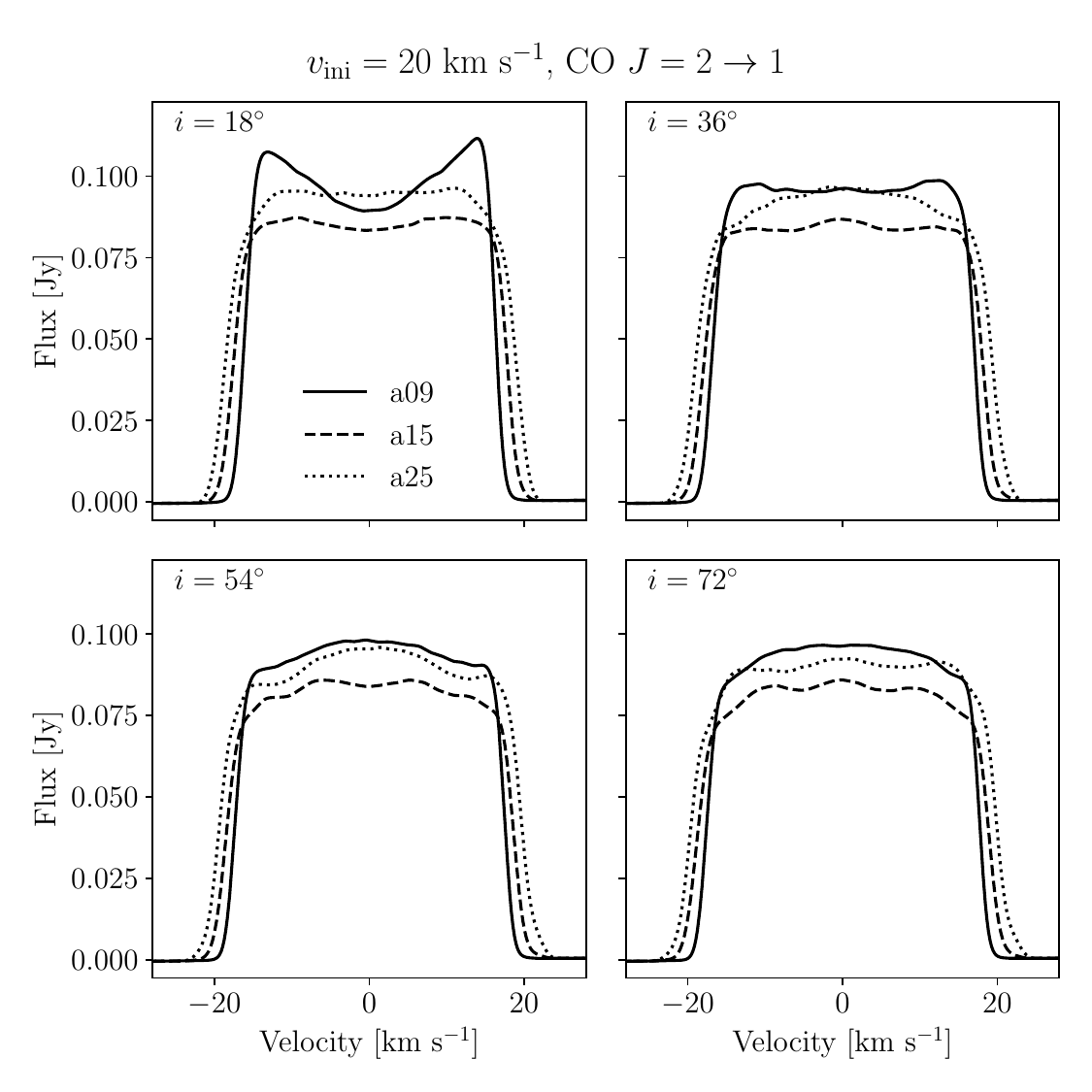}
        \caption{Same as Fig.~\ref{Fig:inclinations_v10}, for the models with initial wind velocities of 20 km s$^{-1}$.}
        \label{Fig:inclinations_v20}
    \end{minipage}
\end{figure}

\newpage
\subsection{Higher S/N}

This section contains an additional figure, visualising the observability of features when assuming a higher S/N value of 20.
\begin{figure}[htbp]
	\includegraphics[width = \columnwidth]{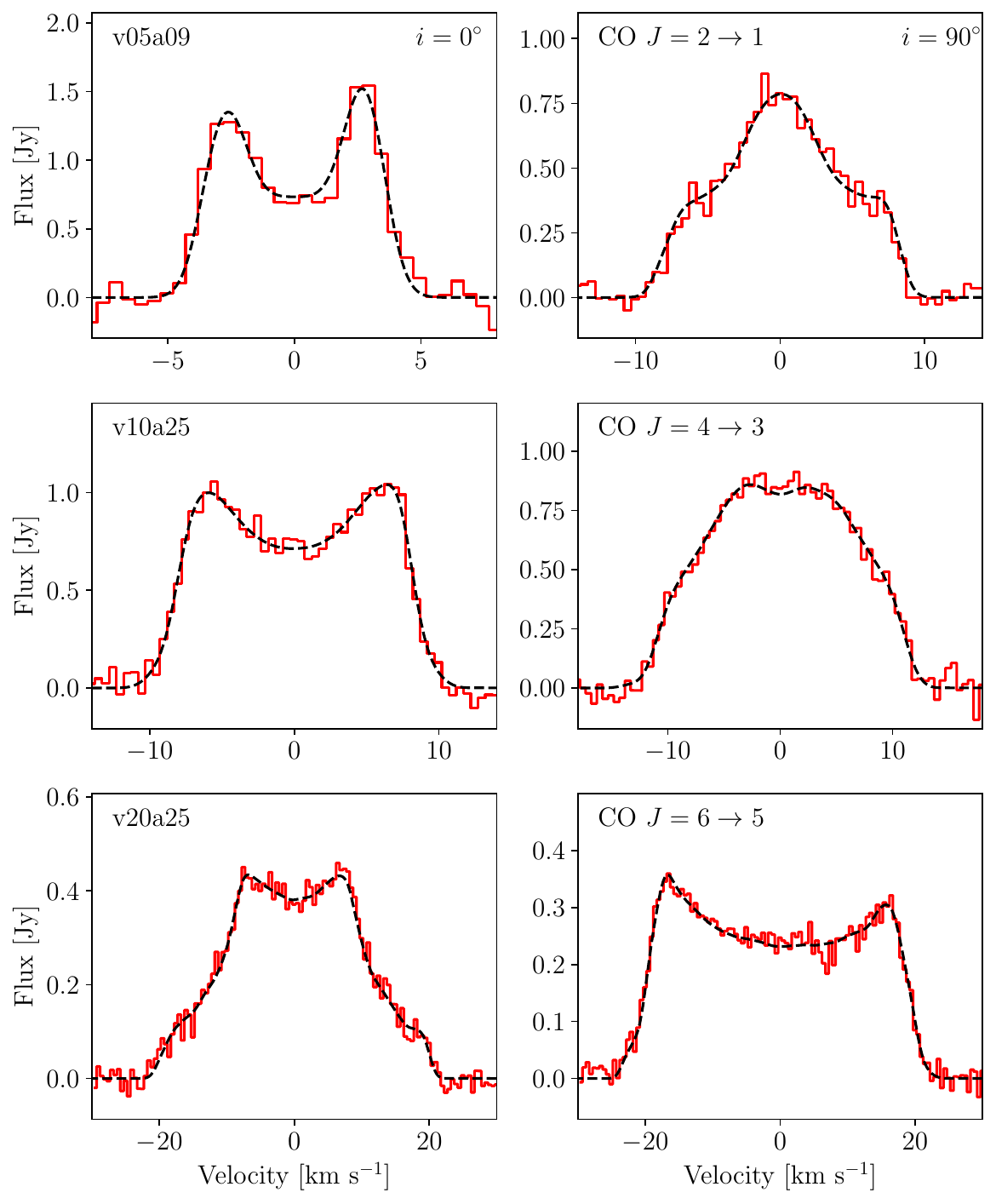}
	\caption{Synthetic CO spectral lines for three selected models. From top to bottom these are v05a09, v10a25, and v20a25 with the $J$~$=$~$2{\rightarrow}1, 4{\rightarrow}3, 6{\rightarrow}5$ transitions, respectively. The spectral lines are created viewing face-on (left column) and edge-on (right column). The black dashed lines show the original spectral line, and the red line has the addition of white noise assuming a S/N of 20, and an increased velocity bin size to 500 m s$^{-1}$.}
    \label{fig:noise_20}
\end{figure}

\end{document}